\documentclass[aps,prl,twocolumn,showpacs,superscriptaddress,preprintnumbers]{revtex4-2}
\usepackage{amsmath}
\usepackage{epsfig}
\usepackage{color}
\usepackage{endnotes}
\usepackage{natbib}
\usepackage[colorlinks,breaklinks]{hyperref}
\usepackage{nicefrac}
\usepackage{bm}
\usepackage{dcolumn}
\usepackage{graphics}
\usepackage{amsmath}
\usepackage{amssymb}
\usepackage{color}
\usepackage{changes}

\newcommand{\bs}[1]{\ensuremath{\boldsymbol{#1}}}

\newcommand{\be}{\begin{equation}}
\newcommand{\ee}{\end{equation}}
\newcommand{\bea}{\begin{align}}
\newcommand{\eea}{\end{align}}
\newcommand{\beqa}{\begin{eqnarray}}
\newcommand{\eeqa}{\end{eqnarray}}

\begin{document}

\preprint{LA-UR-22-32748}

\title{Nuclear three-body short-range correlations in coordinate space}

\author{Ronen Weiss}
\affiliation{Theoretical Division, Los Alamos National Laboratory, Los Alamos, New Mexico 87545, USA}
\author{Stefano Gandolfi}
\affiliation{Theoretical Division, Los Alamos National Laboratory, Los Alamos, New Mexico 87545, USA}

\date{\today}

\begin{abstract} 
We study the effects of three-nucleon short-range correlations on nuclear coordinate-space densities.
For this purpose, novel three-body densities are calculated for ground state nuclei using the auxiliary-field diffusion Monte Carlo method. The results are analyzed in terms of the Generalized Contact Formalism, extended to include three-body correlations, revealing the universal behavior of nucleon triplets at short distances. We identify the quantum numbers of such correlated triplets and extract scaling factors of triplet abundances that can be compared to upcoming inclusive electron scattering data. 
\end{abstract}

\maketitle

Short-range correlations (SRCs) are an integral part of strongly interacting many-body quantum systems, including nuclei and atomic systems. Strong SRCs between nucleons pose one of the main challenges in the description of nuclei.
Accounting for the impact of SRC physics is crucial for the description of two-body densities and momentum distributions \cite{Rios:2009gb,Feldmeier:2011qy,Alvioli:2012qa,Wiringa:2013ala,Weiss:2015mba,Weiss:2016obx}, electron and neutrino scattering \cite{Benhar:1994hw,cda96,Weiss:2018tbu,Pastore:2019urn,Andreoli:2021cxo}, spectroscopic factors \cite{Pandharipande:1997zz,Dickhoff:2004xx,Lapikas:1993uwd,Kramer:2000kc,JeffersonLabHallA:2000dxx,Gade:2008zz,Flavigny:2013bha,Kay:2013bsa,Tostevin:2014usa,Atar:2018dhg,Kawase:2018ojr,Paschalis:2018zkx,Atkinson:2018nvp}, neutrinoless double beta decay matrix elements \cite{Simkovic:2009pp,PhysRevLett.120.202001,Weiss:2021rig}, neutron star properties \cite{hen15,Cai:2015xga,Li:2018lpy,Carbone_2012} and more.

The properties of nuclear SRC pairs, i.e. two nucleons found close together inside a nucleus, have been studied thoroughly in the last decades \cite{Atti:2015eda,Hen:2016kwk,Arrington:2022sov}. The universal features of SRC pairs and the dominance of neutron-proton $(np)$ pairs have been established based on both experimental studies, using mainly large momentum-transfer quasi-elastic electron- and proton-scattering reactions \cite{Frankfurt81,Frankfurt88,frankfurt93,egiyan02,egiyan06,fomin12,Schmookler:2019nvf,JeffersonLabHallA:2020wrr,Li:2022fhh,patsyuk2021unperturbed,tang03,piasetzky06,shneor07,subedi08,korover14,Cohen:2018gzh,hen14,Duer:2018sxh,CLAS:2020rue,schmidt20}, and ab-initio many-body calculations \cite{Feldmeier:2011qy,Alvioli:2012qa,Alvioli:2013qyz,Wiringa:2013ala,Rios:2013zqa,Ryckebusch:2019oya,neff15,ryckebusch15}. A connection to the internal structure of nucleons in nuclei was also revealed \cite{CiofidegliAtti:2007ork,Hen:2016kwk,weinstein11,Hen12,Arrington:2012ax,Hen:2013oha,Chen:2016bde,Schmookler:2019nvf}. 

Unlike two-body SRCs, the features and importance of three-nucleon SRCs are mostly unknown.
There are currently significant efforts to study such correlated triplets experimentally \cite{Arrington_E12-06-105,Ye18,Sargasian19}, but there has been no clear identification of SRC triplets. Similarly, nuclear many-body ab-initio calculations that allow direct access to triplet properties have not been performed so far. Theoretical studies of three-nucleon SRCs are important for guiding the experimental efforts and data analysis and for revealing the properties and impact of such triplets. 

Following the development of the contact theory for zero-range interactions \cite{Tan08a,Tan08b,Tan08c,Braaten12,gandolfi11}, the Generalized Contact Formalism (GCF) was introduced to study nuclear SRCs \cite{Weiss14,Weiss:2015mba,Weiss:2016bxw}. Based on the asymptotic factorization of nuclear wave functions, the GCF allows us to identify the impact of SRC pairs on many different quantities \cite{Weiss:2015mba,Weiss:2016obx,Cruz-Torres2020,Weiss:2018tbu,schmidt20,Pybus:2020itv,Duer:2018sxh,weiss2020inclusive,CLAS:2020rue,patsyuk2021unperturbed,Weiss:2021rig,Weiss:2016bxw,Weiss14,Weiss_EPJA16}, resulting in an overall comprehensive and consistent picture of two-nucleon SRCs.

In this work we extend the GCF to include the description of three-nucleon SRCs. We first discuss the expected factorized form of the many-body wave function when three nucleons are found close to each other and the possible quantum numbers of such triplets. We then define three-body nuclear contacts, describing the abundance of SRC triplets, and derive an asymptotic expression for the three-body nuclear density. Finally, we present quantum Monte Carlo calculations of such densities, verify the predictions of the GCF and extract scaling factors of triplet abundances that are connected to inclusive electron-scattering cross sections. 

When two nucleons are close to each other in a nucleus with $A$ nucleons, the many-body wave function $\Psi$ factorizes to a two-body part, describing the correlated pair, and a function describing the rest of the nucleons in the system \cite{Weiss:2015mba}
\be
\Psi \xrightarrow[r_{ij}\rightarrow 0]{}
\sum_\alpha \varphi^\alpha(\bs{r}_{ij}) A^\alpha(\bs{R}_{ij},\{\bs{r}_k\}_{k\neq i,j}).
\ee
This is the GCF factorization. Here, $\bs{r}_k$ denotes the single-nucleon coordinate, $\bs{r}_{ij}\equiv \bs{r}_j-\bs{r}_i$ and $\bs{R}_{ij}\equiv (\bs{r}_i+\bs{r}_j)/2$ are the relative and center-of-mass (CM) coordinates, and $\alpha$ denotes the quantum numbers of the pair.
$\varphi^\alpha(\bs{r}_{ij})$ describes the dynamics of the correlated pair and is defined as the solution of the zero-energy two-body Schr\"odinger equation. As such it is universal, i.e. nucleus-independent, but depends on the nucleon-nucleon interaction model. This factorization was verified using ab-initio calculations \cite{Weiss:2015mba,Weiss:2016obx,Cruz-Torres2020,Alvioli:2016wwp}, and is supported by renormalization-group arguments \cite{Anderson2010,Bogner12,Tropiano2021} and the Coupled-Cluster expansion \cite{Beck_CC}.

The above two-body factorization is valid when none of the remaining $A-2$ nucleons is close to the correlated pair. If one nucleon is close enough to such a pair, i.e. when three nucleons are close to each other,
we expect the many-body nuclear wave function to factorize in the following way
\be \label{eq:full_asymp_3N}
{\Psi} \xrightarrow[x_{ij},x_{ijl}\rightarrow 0]{}
\sum_\beta {\varphi}_{ijl}^\beta (\bs{{x}}_{ij},\bs{{x}}_{ijl})
 {B}_{ijl}^\beta (\bs{R}_{ijl},\{\bs{r}_m\}_{m\neq i,j,l}).
\ee
We used here the Jacobi coordinates $\bs{x}_{ij}\equiv\bs{r}_j-\bs{r}_i$ and $\bs{x}_{ijl}\equiv \bs{r}_l-(\bs{r}_i+\bs{r}_j)/2$ and the triplet CM coordinate $\bs{R}_{ijl} \equiv(\bs{r}_i+\bs{r}_j+\bs{r}_l)/3$.
${\varphi}_{ijl}^\beta$ describes the dynamics of the SRC triplet and is defined as a zero-energy solution of the three-body Schr\"odinger equation with quantum numbers given by $\beta$ (with the same nuclear-interaction model used to define $\Psi$). We note that there could be multiple independent zero-energy solutions with the same quantum numbers due to different possible boundary conditions. 
We assume here that a single three-body wave function per channel (i.e. per quantum numbers $\beta$) is sufficient, similar to the case of two-nucleon SRCs \cite{Weiss17_CoupledChannels, Weiss18_CoupledChannels}, and we provide below numerical evidence for this claim based on ab-initio calculations. The connection of three-nucleon SRCs to the zero-energy three-body eigenstates is discussed also in Ref. \cite{Beck_CC}.

For realistic nuclear interactions, each channel $\beta$ is defined by the quantum numbers
$\beta = (\pi_\beta,j_\beta,m_\beta,t_\beta,t_{z,\beta})$,
where $\pi_\beta$ is the parity, $j_\beta$ and $m_\beta$ are the total angular momentum and its projection, and $t_\beta$ and $t_{z,\beta}$ are the total isospin of the triplet and its projection.
At short distances, we expect to see a dominant contribution of channels that include a zero angular-momentum component $\ell=0$. This is possible only for channels with positive parity $\pi=+$. Since the total spin of three nucleons is either $s=1/2$ or $s=3/2$, the dominant channels should have $j=1/2$ or $j=3/2$.
Isospin value of $t=3/2$ is suppressed due to Pauli blocking, as three nucleons cannot be in the same location with $t=3/2$. Therefore, $t=1/2$ is expected to be dominant at short distances. Similarly, the symmetric $s=3/2$ component is also suppressed due to Pauli blocking. Thus, $s=1/2$ and therefore $j=1/2$ is expected to be dominant at short distances. To conclude, among the infinite number of three-body channels in Eq. \eqref{eq:full_asymp_3N}, we expect the dominant channels to be with the quantum number $\pi=+,j=1/2,t=1/2$ (and $m=\pm1/2$ and $t_z=\pm1/2$). These channels correspond to the quantum numbers of $^3$He and $^3$H ground states. It means also that proton-proton-proton $(ppp)$ and neutron-neutron-neutron $(nnn)$ triplets are expected to be suppressed at short distances compared to proton-proton-neutron $(ppn)$ and proton-neutron-neutron $(pnn)$ triplets. Notice that while the $np$ dominance of two-body SRCs is caused by the tensor force \cite{Alvioli:2007zz,schiavilla07}, here it is the Pauli principle that leads to the $t=1/2$ dominance for three-body SRCs. See more details regarding the structure of ${\varphi}_{ijl}^\beta$ and the dominant channels in the supplementary materials.

Based on Eq. \eqref{eq:full_asymp_3N}, we can now define the three-nucleon contact matrix
\be \label{eq:3N_JM_contacts}
C_3^{\beta \gamma}(JM) = 
\frac{A(A-1)(A-2)}{6} \langle B_{123}^\beta | B_{123}^\gamma \rangle.
\ee 
$J$ and $M$ above are the total angular momentum and projection of $\Psi$. The combinatorical factor is suitable assuming $\Psi$ is fully anti-symmetric. Three-body contacts were similarly defined for the zero-range limit \cite{Braaten11,Werner12_B} and for Helium atoms \cite{Bazak20}.
Based on orthogonality properties of the $B_{ijl}^\beta$ functions, we can conclude that if $m_\beta\neq m_\gamma$, then $C_3^{\beta \gamma}(JM)=0$. Similarly, if $t_{z,\beta}\neq t_{z,\gamma}$ or $\pi_\beta \neq \pi_\gamma$, then $C_3^{\beta \gamma}(JM)=0$ as well, assuming $\Psi$ has well defined isospin $T$ and parity. 
This is generally not the case for $j$ or $t$. Nevertheless, for $J=0$ nuclei, we do get $C_3^{\beta\gamma}(00)=0$ if $j_\beta \neq j_\gamma$, and for $T=0$, we get $C_3^{\beta\gamma}(JM)=0$ if $t_\beta \neq t_\gamma$.
We can also show that the sum over $m_\beta$ of diagonal contacts $\sum_{m_\beta}C_3^{\beta\beta}(JM)$ is independent of $M$. Derivations appear in the supplementary materials.

It is useful to define also the $M$-averaged three-body nuclear contacts
\be\label{eq:3N_ave_contacts}
C_3^{\beta\gamma}=\frac{1}{2J+1}\sum_M C^{\beta\gamma}(JM).
\ee
The averaged contacts are diagonal in both $j$ and $m$, {\it i.e.} $C_3^{\beta \gamma}=0$ if $m_\beta\neq m_\gamma$ or $j_\beta \neq j_\gamma$, and 
they are also independent of $m_\beta$ and $m_\gamma$. See derivations in the supplementary materials. The averaged contacts inherit the properties of 
$C_3^{\beta\gamma}(JM)$ regarding parity and the isospin projection $t_z$. Notice that the averaged contacts are still generally not diagonal in $t_\beta$.

To study the implications of SRC triplets, we consider in this work three-body densities in coordinate space. Specifically, we consider the three-body density describing the probability of finding three nucleons inside a nucleus in a triangle with sides of length $r_{12}$, $r_{13}$ and $r_{23}$
\begin{align} \label{eq:3N_density}
& \rho_3(r_{12},r_{13},r_{23}) = \frac{A(A-1)(A-2)}{6}
\nonumber \\ &\times  \langle \Psi | \delta(|\bs{r}_1-\bs{r}_2|-r_{12})
\delta(|\bs{r}_1-\bs{r}_3|-r_{13})\delta(|\bs{r}_2-\bs{r}_3|-r_{23}) | \Psi \rangle.
\end{align}
In the limit of $r_{12},r_{13},r_{23} \rightarrow 0$, we can use Eqs. \eqref{eq:full_asymp_3N}-\eqref{eq:3N_ave_contacts} to obtain
\begin{align} \label{eq:3N_density_asymptotic}
& \rho_3(r_{12},r_{13},r_{23}) \xrightarrow[r_{12},r_{13},r_{23}\rightarrow 0]{}
\sum_{\pi_\beta,j_\beta,t_\beta}
\left(\sum_{m_\beta,t_{z,\beta}} C_3^{\beta}\right)
 \langle \varphi_{123}^\beta |
\nonumber \\ &\times   \delta(|\bs{r}_1-\bs{r}_2|-r_{12})
\delta(|\bs{r}_1-\bs{r}_3|-r_{13})\delta(|\bs{r}_2-\bs{r}_3|-r_{23}) | \varphi_{123}^\beta \rangle.
\end{align}
$\rho_3(r_{12},r_{13},r_{23})$ is independent of $M$ and therefore we could use here the M-averaged contacts, which are diagonal in $\pi_\beta,j_\beta,m_\beta,t_{z,\beta}$. In addition, the three-body part in Eq. \eqref{eq:3N_density_asymptotic} is diagonal in $t_\beta$ and, therefore, we are left only with the diagonal contacts $ C_3^{\beta}\equiv C_3^{\beta\beta}$. Finally, the three-body part is also independent of $m_\beta$ and $t_{z,\beta}$.
Eq. \eqref{eq:3N_density_asymptotic} reveals the interpretation of the diagonal three-body contacts $ C_3^{\beta}$ - they measure the probability of finding three-nucleons close together with quantum numbers $\beta$.

$\pi_\beta,j_\beta$ and $t_\beta$ are the quantum numbers governing the $r$-dependence of $\rho_3$ at short distances. As discussed before, a single set of values with $\pi_\beta=+,j_\beta=1/2,t_\beta=1/2$ should be dominant. Therefore, we expect to find a universal behavior of $\rho_3$ at short-distances for all nuclei, i.e. the same $r$-dependence with only a global scaling factor that depends on the nucleus.
We can also consider isospin-projected densities $\rho_3^t(r_{12},r_{13},r_{23})$, 
by inserting the appropriate three-body projection operator in Eq. \eqref{eq:3N_density}. For these quantities, we expect to see a dominance of $t=1/2$ over $t=3/2$ as discussed above.

In order to verify these GCF predictions regarding three-nucleons SRCs, we now turn to ab-initio calculations of $\rho_3(r_{12},r_{13},r_{23})$ for the ground-state nuclei $^3$He, $^4$He, $^6$Li and $^{16}$O. We used the auxiliary-field diffusion Monte Carlo (AFDMC) method \cite{Carlson:2014vla,Lonardoni:2018prc} combined with the N2LO local chiral interaction with the E1 parametrization of the three-body force \cite{Gezerlis:2013ipa,Gezerlis:2014,Lynn:2016,Lonardoni:2018prc}. We focus here on the $R_0=1.0$ fm cutoff but show also some results for $R_0=1.2$ fm. 

\begin{figure}\begin{center}
\includegraphics[width=\linewidth]{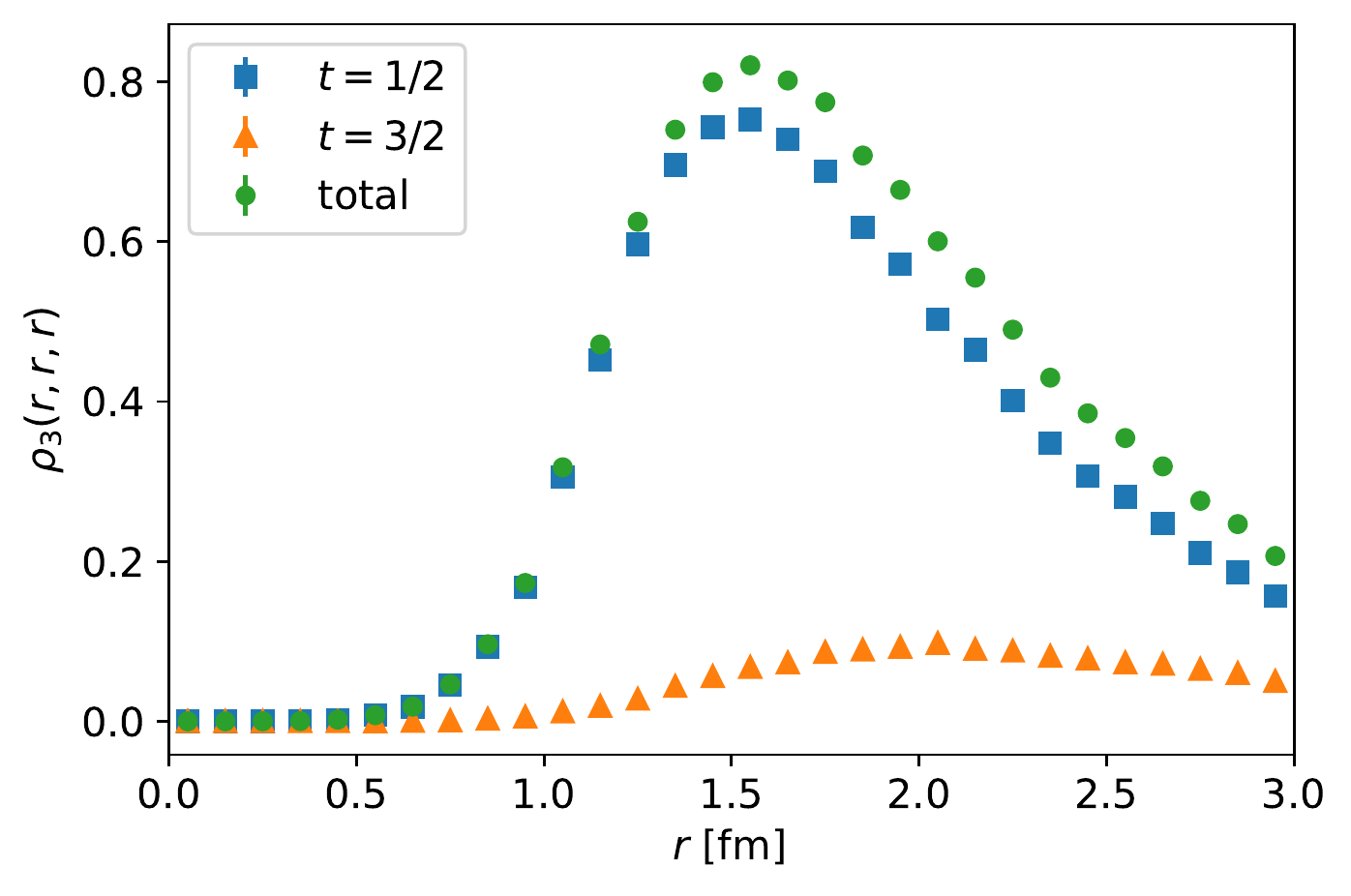}
\caption{\label{fig:equi_6Li_T_proj}
$^{6}$Li three-body density for equilateral triangles as a function of the triangle side using the AFDMC method and the N2LO(1.0) interaction. Projections to $t=1/2$ and $t=3/2$ triplets are shown together with the total density.}
\end{center}\end{figure}

We first start with investigating $\rho_3^t$ in order to compare the $t=1/2$ and $t=3/2$ densities.
The three-body density for $^{6}$Li is shown in Fig. \ref{fig:equi_6Li_T_proj} for the equilateral triangle, i.e. $\rho_3(r,r,r)$ as a function of $r$. We can clearly see that at short distances the $t=1/2$ component is dominant. As $r$ increases the contribution of $t=3/2$ triplets grows. We note that the total number of $t=1/2$ and $t=3/2$ triplets in $^{6}$Li is $16$ and $4$ respectively. The contribution of $t=1/2$ triplets in $^{6}$Li for $r<1$ fm is larger than $95\%$ of all triplets, i.e. significantly larger than the total combinatorial part of such triplets which is $80\%$. $t=1/2$ dominance is seen also for other geometries, e.g. isosceles triangles, and also for $^{16}$O ($^3$He and $^4$He include only $t=1/2$ triplets).
These results agree with our expectation due to the Pauli exclusion rule.

\begin{figure} \begin{center}  
    \begin{tikzpicture}
        \node[anchor=south west,inner sep=0] (image) at (0,0) {\includegraphics[width=\linewidth]
{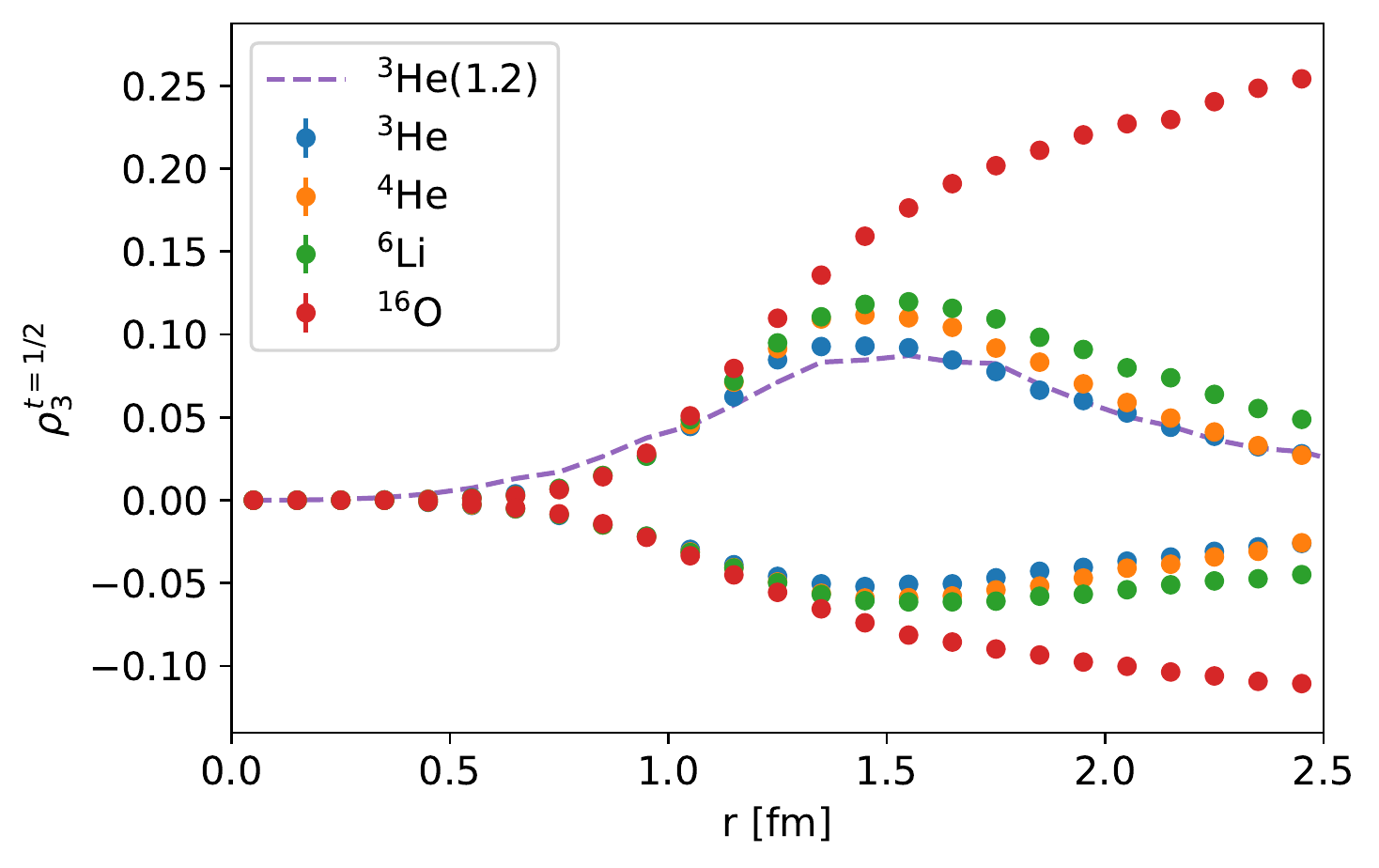}};
        \begin{scope}[x={(image.south east)},y={(image.north west)}]
            \node[anchor=south west,inner sep=0] (image) at (0.455,0.72) {\includegraphics[width=1.4cm,trim=14.1cm 8.4cm 16.3cm 7.7cm,clip]{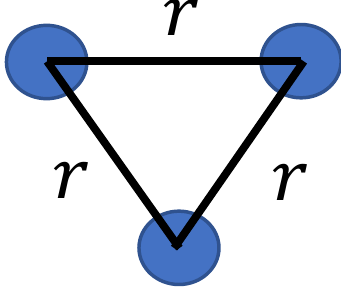}};
            \node[anchor=south west,inner sep=0] (image) at (0.23,0.17) {\includegraphics[width=1.7cm,trim=15.5cm 7cm 13.5cm 8.7cm,clip]{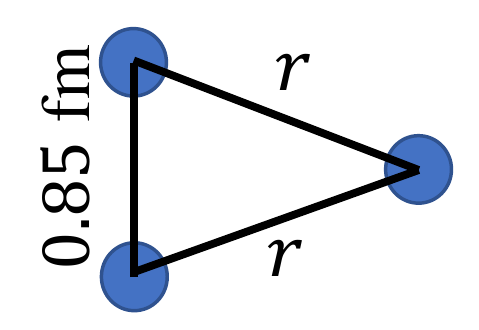}};  
        \end{scope}
    \end{tikzpicture}
    \caption{\label{fig:THalf_scaled} 
AFDMC $\rho_3^{t=1/2}$ densities for $^3$He, $^4$He, $^6$Li and $^{16}$O for both equilateral triangle and isosceles triangle using the N2LO(1.0) interaction (circles). For the latter, the base is fixed at length of $0.85$ fm and the densities are multiplied by minus one (see text for details). The $^4$He, $^6$Li and $^{16}$O densities are each multiplied by a scaling factor (the same factor for both geometries). $^3$He equilateral-triangle density is shown also for the N2LO(1.2) interaction (dashed line).
}
\end{center} \end{figure}

We can now focus on the $t=1/2$ component and compare the behavior of different nuclei. 
We present in Fig. \ref{fig:THalf_scaled} the density $\rho_3^{t=1/2}$ for all available nuclei for both equilateral and isosceles triangle geometries using the N2LO(1.0) interaction. For the isosceles triangle, we fix the base to be of length $a=0.85$ fm. 
The $^4$He, $^6$Li and $^{16}$O calculations are rescaled so that their shape at short distances can be compared to $^3$He. Only for plotting purposes, the densities of the isosceles triangle are multiplied by minus one to separate them from the equilateral-triangle results. We can see that, for each of the geometries, the $r$-dependence of $\rho_3^{t=1/2}$ is the same at short distances ($r \lesssim 1.1$ fm) for all nuclei, as all densities coincide with the $^3$He density.
This shows the universal behavior of SRC triplets as predicted by the GCF. Indeed, a single $t=1/2$ channel is dominant here (otherwise the densities would not coincide) due to the dominance of $\ell=0$ at short distances. Also, as assumed in Eq. \eqref{eq:full_asymp_3N}, a single three-body wave function for the dominant quantum numbers is sufficient as all nuclei behave like the bound $^3$He.
It should be emphasized that the same scaling factor is applied to both the equilateral and isosceles cases for each nucleus, in agreement with Eq. \eqref{eq:3N_density_asymptotic}. The same behavior is seen for other configurations involving three particles close together, e.g. other isosceles triangles as long as the base and legs lengths are small. This result is an important validation of the asymptotic three-body factorization of the many-body wave function, Eq. \eqref{eq:full_asymp_3N}.
We also include in Fig. \ref{fig:THalf_scaled} the $^3$He equilateral-triangle density using the N2LO(1.2) interaction. We can see that the short-distance behavior in this case is different. This shows that the three-body wave functions of the GCF ${\varphi}_{ijl}^\beta$ indeed depend on the model of the interaction.

\begin{figure} \begin{center}  
    \begin{tikzpicture}
        \node[anchor=south west,inner sep=0] (image) at (0,0) {\includegraphics[width=\linewidth]
{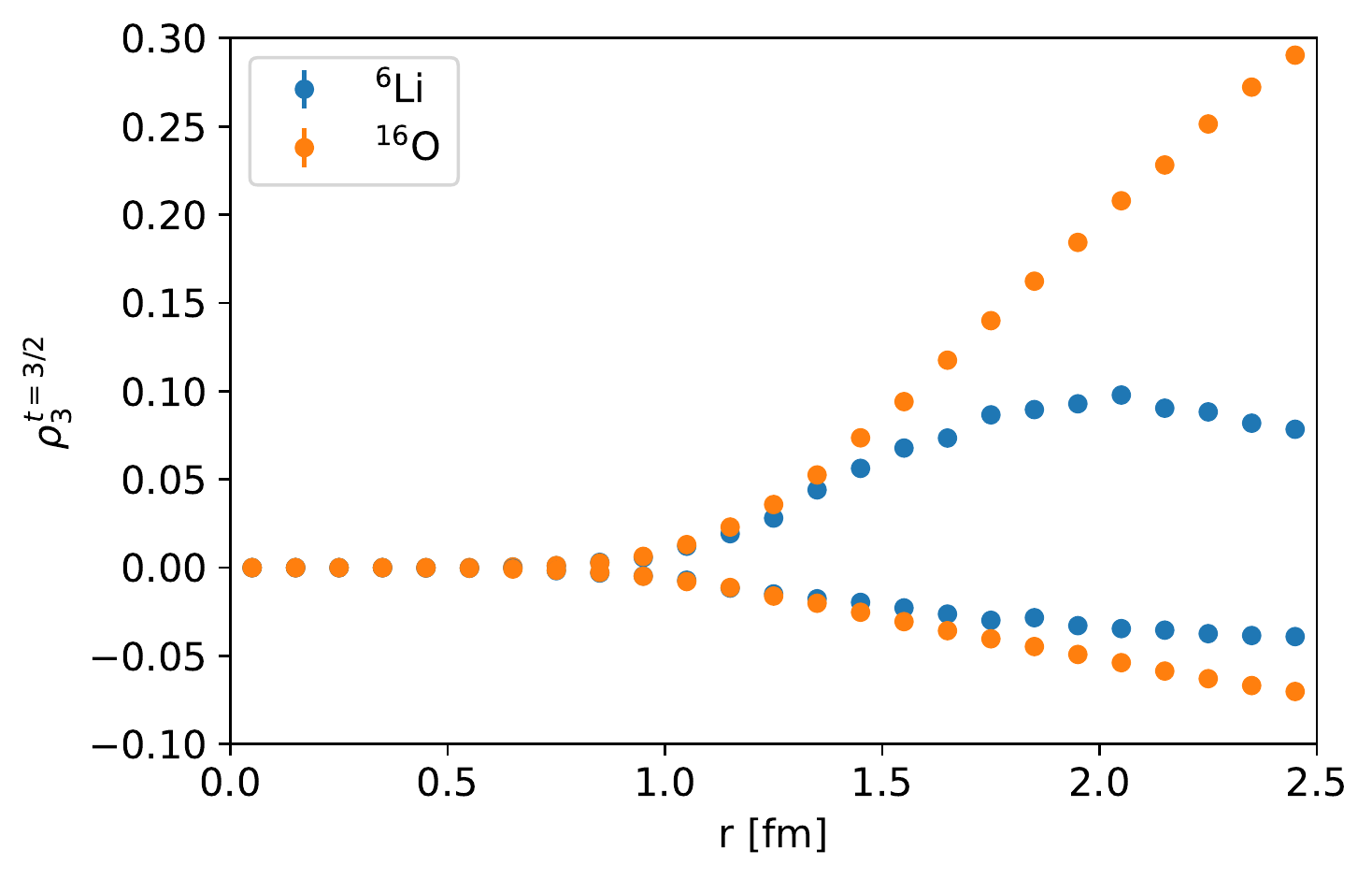}};
        \begin{scope}[x={(image.south east)},y={(image.north west)}]
            \node[anchor=south west,inner sep=0] (image) at (0.4,0.5) {\includegraphics[width=1.7cm,trim=14cm 8cm 16cm 7.5cm,clip]{equilateral_draw.pdf}};
            \node[anchor=south west,inner sep=0] (image) at (0.27,0.156) {\includegraphics[width=1.45cm,trim=15.5cm 7cm 13.5cm 8.7cm,clip]{isosceles_0_85fm_rotated.pdf}};  
        \end{scope}
    \end{tikzpicture}
    \caption{\label{fig:T3Half_scaled} 
The same as Fig. \ref{fig:THalf_scaled}, but for the $t=3/2$ component. $^{16}$O density is rescaled.
}
\end{center} \end{figure}

Despite the dominance of the $t=1/2$ component, we can also look into the properties of the $t=3/2$ density. We note that among the nuclei considered in this work, only $^6$Li and $^{16}$O contain a $t=3/2$ three-body component.
The equilateral-triangle and isosceles-triangle $\rho_3^{t=3/2}$ densities are shown in Fig. \ref{fig:T3Half_scaled} with rescaled $^{16}$O densities, similar to Fig. \ref{fig:THalf_scaled}. We can see that also for this case a universal short-distance behavior exists. This indicates that the asymptotic factorization holds for $t=3/2$ triplets and that there is a single dominant $t=3/2$ channel. The latter is expected to have the quantum numbers $\pi=+$ and $j=1/2$. Calculations for additional nuclei with $t=3/2$ component are needed to validate these conclusions.

The scaling factors used in Figs. \ref{fig:THalf_scaled} and \ref{fig:T3Half_scaled} are equal to a sum of three-nucleon contact values (relative to $^3$He). For $t=1/2$, it is the sum of contacts 
\be
C_3^{t=1/2}\equiv \sum_{m,t_z} C_3^{\beta_{1/2}},
\ee
where $\beta_{1/2}$ denotes the quantum numbers of the leading $t=1/2$ three-body channel.
Notice that both $ppn$ and $pnn$ triplets (i.e. $t_z=\pm 1/2$) are included here. 
As mentioned above, such contact values are proportional to the probability of finding correlated triplets in a given nucleus. Similar probabilities can be accessible using large momentum transfer quasi-elastic inclusive electron-scattering experiments. For the case of two-body SRCs, pair abundances were extracted from inclusive measurements at appropriate kinematics, defining $a_2$ as the per-nucleon cross section ratio with respect to the deuteron \cite{Frankfurt81,Frankfurt88,frankfurt93,egiyan02,egiyan06,fomin12,Schmookler:2019nvf,JeffersonLabHallA:2020wrr,Li:2022fhh}. 
There are ongoing experimental efforts to extract such three-nucleon abundances \cite{Arrington_E12-06-105}. 
 In this case, a cross section ratio can be defined as \cite{Sargasian19}
\be
a_3(A,Z) = \frac{3}{A} \frac{\sigma_{eA}}{(\sigma_{e ^3\textrm{He}}+\sigma_{e ^3\textrm{H}})/2},
\ee
where $\sigma_{eA}$ is the inclusive electron-scattering cross section off nucleus A (with $A$ nucleons, $Z$ protons) at kinematics dominated by three-body SRCs. Interpreting $a_3$ as the ratio of three-nucleon SRC abundances, it is connected to the nuclear contacts for a symmetric nucleus $A$ by
\be
a_3(A,Z) = \frac{3}{A} \frac{C_3^{t=1/2}(A)}{C_3^{t=1/2}(^3\textrm{He})}.
\ee
We consider here only the leading contribution of $t=1/2$ triplets.
It should be noted that, similar to the case of two-body SRCs \cite{weiss2020inclusive}, different effects can influence this interpretation of $a_3$, such as the CM motion of triplets, excitation energy of the $A-3$ system and contribution of $t=3/2$ triplets. In addition, the dependence of $a_3$ on kinematic variables (such as x-bjorken and $Q^2$) should cancel in the ratio if we consider a symmetric nucleus in the numerator and the sum of $^3$He and $^3$H in the denominator (ignoring the above effects), due to similar contribution of $ppn$ and $nnp$ triplets in the numerator and denominator. For the case of non-symmetric nuclei or if using only the $^3$He cross section in the denominator, a more careful analysis of the reaction is needed.

Contact ratios extracted from the AFDMC calculations are presented in Table \ref{tab:fitted_contacts}.
Their values were fitted to the equilateral three-body density based on Eq. \eqref{eq:3N_density_asymptotic}, where only the leading channel (for each value of $t$) is considered. Since we are looking on contact ratios, there is no need to calculate the functions $\varphi_{ijl}^\beta$. Uncertainties were estimated by varying the lower and upper limits of the fitting range between $0.1-0.4$ fm and $1-1.2$ fm, respectively.
The $t=1/2$ contact ratio provides a prediction for the value of $a_3$ for $^4$He, $^6$Li and $^{16}$O. 
We can see that the per-nucleon $t=1/2$ ratios for $^{4}$He and $^{16}$O are similar, consistent with a $^4$He-cluster structure of $^{16}$O. It is also interesting to note that the per-nucleon $t=1/2$ ratio for $^6$Li is smaller than that of $^4$He. Results for additional nuclei are needed in order study the $A$-dependence of $a_3$.

In a recent work \cite{Sargasian19,Sargsian:2012sm,Sargsian_private}, Sargsian {\it et. al.} suggested a connection between two-body and three-body abundances $a_3(A)=1.12 a_2(A)^2/a_2(^3\mathrm{He})$, leading to a value of $a_3(^4\mathrm{He}) \approx 3.15$ based on the experimental values of $a_2$ from Ref. \cite{Arrington:2012ax}. This is smaller than the value we obtained here (table \ref{tab:fitted_contacts}).
We emphasize that in the GCF approach three-nucleon abundances are generally independent of two-nucleon abundances. Inclusive experimental measurements might be able to clear this issue.

\begin{table}
\begin{tabular}{c  c  c  c }
  \hline
\hline                       
contact ratio  & $A=4$ & $A=6$ & $A=16$  \\
\hline
  $t=1/2$ & $3.8\pm 0.3$ & $3.1\pm 0.3$ & $4.2\pm0.5$ \\
  $t=3/2$ & -  & - & $5.4\pm0.8$\\
\hline
\hline    
\end{tabular}
 \caption{\label{tab:fitted_contacts} 
Per nucleon three-body contact ratio for $t=1/2$ (with respect to $^3$He) and for $t=3/2$
(with respect to $^6$Li), i.e.
$\frac{3}{A}\frac{C_3^{t=1/2}(A)}{C_3^{t=1/2}(^3\textrm{He})}$, and
$\frac{6}{A}\frac{C_3^{t=3/2}(A)}{C_3^{t=3/2}(^6\textrm{Li})}$, respectively.
}
\end{table}

To summarize, we have studied here the properties of three-nucleon SRCs in coordinate space.
The GCF was extended to include correlated triplets, described as an isolated and universal subsystem within the many-body nucleus. The leading $t=1/2$ channel was identified and three-nucleon contacts were defined. Using novel AFDMC ab-initio calculations of three-body densities, the $t=1/2$ dominance and universality of such triplets at short-distances was established numerically. Specifically, we found that three nucleons at short distances behave like the bound $^3$He wave function. We have also extracted the values of the leading $t=1/2$ and $t=3/2$ contact ratios, describing the scaling of SRC triplet abundances. The connection to inclusive electron-scattering cross sections was discussed.

This work opens the path for additional studies of SRC triplets, including their impact on two-body densities, momentum distributions, spectral functions, electron and neutrino scattering off nuclei and neutrinoless double beta decay matrix elements. The sensitivity of SRC triplet properties to the nuclear interaction model and specifically to the three-nucleon force should also be further studied. This is also an important step towards a systematic short-range expansion of the nuclear wave function.

We would like to thank J. Carlson, J. Martin, S. Novario and I. Tews, 
for helpful discussions.
The work of R.W. was supported by the Laboratory Directed Research and Development program of Los Alamos National Laboratory under project number 20210763PRD1.
The work of S. G. was supported by U.S. Department of Energy, Office of Science,
Office of Nuclear Physics, under Contract No. DE-AC52-06NA25396,
by the DOE NUCLEI SciDAC Program, and by the DOE Early Career Research Program. 
Computer time was provided by the Los Alamos National Laboratory Institutional Computing Program, which is supported by the U.S. Department of Energy National Nuclear Security Administration under Contract No. 89233218CNA000001.

\bibliography{references}

\begin{thebibliography}{98}%
\makeatletter
\providecommand \@ifxundefined [1]{%
 \@ifx{#1\undefined}
}%
\providecommand \@ifnum [1]{%
 \ifnum #1\expandafter \@firstoftwo
 \else \expandafter \@secondoftwo
 \fi
}%
\providecommand \@ifx [1]{%
 \ifx #1\expandafter \@firstoftwo
 \else \expandafter \@secondoftwo
 \fi
}%
\providecommand \natexlab [1]{#1}%
\providecommand \enquote  [1]{``#1''}%
\providecommand \bibnamefont  [1]{#1}%
\providecommand \bibfnamefont [1]{#1}%
\providecommand \citenamefont [1]{#1}%
\providecommand \href@noop [0]{\@secondoftwo}%
\providecommand \href [0]{\begingroup \@sanitize@url \@href}%
\providecommand \@href[1]{\@@startlink{#1}\@@href}%
\providecommand \@@href[1]{\endgroup#1\@@endlink}%
\providecommand \@sanitize@url [0]{\catcode `\\12\catcode `\$12\catcode
  `\&12\catcode `\#12\catcode `\^12\catcode `\_12\catcode `\%12\relax}%
\providecommand \@@startlink[1]{}%
\providecommand \@@endlink[0]{}%
\providecommand \url  [0]{\begingroup\@sanitize@url \@url }%
\providecommand \@url [1]{\endgroup\@href {#1}{\urlprefix }}%
\providecommand \urlprefix  [0]{URL }%
\providecommand \Eprint [0]{\href }%
\providecommand \doibase [0]{https://doi.org/}%
\providecommand \selectlanguage [0]{\@gobble}%
\providecommand \bibinfo  [0]{\@secondoftwo}%
\providecommand \bibfield  [0]{\@secondoftwo}%
\providecommand \translation [1]{[#1]}%
\providecommand \BibitemOpen [0]{}%
\providecommand \bibitemStop [0]{}%
\providecommand \bibitemNoStop [0]{.\EOS\space}%
\providecommand \EOS [0]{\spacefactor3000\relax}%
\providecommand \BibitemShut  [1]{\csname bibitem#1\endcsname}%
\let\auto@bib@innerbib\@empty
\bibitem [{\citenamefont {Rios}\ \emph {et~al.}(2009)\citenamefont {Rios},
  \citenamefont {Polls},\ and\ \citenamefont {Dickhoff}}]{Rios:2009gb}%
  \BibitemOpen
  \bibfield  {author} {\bibinfo {author} {\bibfnamefont {A.}~\bibnamefont
  {Rios}}, \bibinfo {author} {\bibfnamefont {A.}~\bibnamefont {Polls}},\ and\
  \bibinfo {author} {\bibfnamefont {W.~H.}\ \bibnamefont {Dickhoff}},\
  }\bibfield  {title} {\bibinfo {title} {{Depletion of the nuclear Fermi
  sea}},\ }\href {https://doi.org/10.1103/PhysRevC.79.064308} {\bibfield
  {journal} {\bibinfo  {journal} {Phys. Rev.}\ }\textbf {\bibinfo {volume}
  {C79}},\ \bibinfo {pages} {064308} (\bibinfo {year} {2009})},\ \Eprint
  {https://arxiv.org/abs/0904.2183} {arXiv:0904.2183 [nucl-th]} \BibitemShut
  {NoStop}%
\bibitem [{\citenamefont {Feldmeier}\ \emph {et~al.}(2011)\citenamefont
  {Feldmeier}, \citenamefont {Horiuchi}, \citenamefont {Neff},\ and\
  \citenamefont {Suzuki}}]{Feldmeier:2011qy}%
  \BibitemOpen
  \bibfield  {author} {\bibinfo {author} {\bibfnamefont {H.}~\bibnamefont
  {Feldmeier}}, \bibinfo {author} {\bibfnamefont {W.}~\bibnamefont {Horiuchi}},
  \bibinfo {author} {\bibfnamefont {T.}~\bibnamefont {Neff}},\ and\ \bibinfo
  {author} {\bibfnamefont {Y.}~\bibnamefont {Suzuki}},\ }\bibfield  {title}
  {\bibinfo {title} {{Universality of short-range nucleon-nucleon
  correlations}},\ }\href {https://doi.org/10.1103/PhysRevC.84.054003}
  {\bibfield  {journal} {\bibinfo  {journal} {Phys. Rev. C}\ }\textbf {\bibinfo
  {volume} {84}},\ \bibinfo {pages} {054003} (\bibinfo {year} {2011})},\
  \Eprint {https://arxiv.org/abs/1107.4956} {arXiv:1107.4956 [nucl-th]}
  \BibitemShut {NoStop}%
\bibitem [{\citenamefont {Alvioli}\ \emph
  {et~al.}(2013{\natexlab{a}})\citenamefont {Alvioli}, \citenamefont
  {Ciofi~degli Atti}, \citenamefont {Kaptari}, \citenamefont {Mezzetti},\ and\
  \citenamefont {Morita}}]{Alvioli:2012qa}%
  \BibitemOpen
  \bibfield  {author} {\bibinfo {author} {\bibfnamefont {M.}~\bibnamefont
  {Alvioli}}, \bibinfo {author} {\bibfnamefont {C.}~\bibnamefont {Ciofi~degli
  Atti}}, \bibinfo {author} {\bibfnamefont {L.~P.}\ \bibnamefont {Kaptari}},
  \bibinfo {author} {\bibfnamefont {C.~B.}\ \bibnamefont {Mezzetti}},\ and\
  \bibinfo {author} {\bibfnamefont {H.}~\bibnamefont {Morita}},\ }\bibfield
  {title} {\bibinfo {title} {{Nucleon momentum distributions, their
  spin-isospin dependence and short-range correlations}},\ }\href
  {https://doi.org/10.1103/PhysRevC.87.034603} {\bibfield  {journal} {\bibinfo
  {journal} {Phys. Rev.}\ }\textbf {\bibinfo {volume} {C87}},\ \bibinfo {pages}
  {034603} (\bibinfo {year} {2013}{\natexlab{a}})},\ \Eprint
  {https://arxiv.org/abs/1211.0134} {arXiv:1211.0134 [nucl-th]} \BibitemShut
  {NoStop}%
\bibitem [{\citenamefont {Wiringa}\ \emph {et~al.}(2014)\citenamefont
  {Wiringa}, \citenamefont {Schiavilla}, \citenamefont {Pieper},\ and\
  \citenamefont {Carlson}}]{Wiringa:2013ala}%
  \BibitemOpen
  \bibfield  {author} {\bibinfo {author} {\bibfnamefont {R.~B.}\ \bibnamefont
  {Wiringa}}, \bibinfo {author} {\bibfnamefont {R.}~\bibnamefont {Schiavilla}},
  \bibinfo {author} {\bibfnamefont {S.~C.}\ \bibnamefont {Pieper}},\ and\
  \bibinfo {author} {\bibfnamefont {J.}~\bibnamefont {Carlson}},\ }\bibfield
  {title} {\bibinfo {title} {{Nucleon and nucleon-pair momentum distributions
  in $A \le 12$ nuclei}},\ }\href {https://doi.org/10.1103/PhysRevC.89.024305}
  {\bibfield  {journal} {\bibinfo  {journal} {Phys. Rev. C}\ }\textbf {\bibinfo
  {volume} {89}},\ \bibinfo {pages} {024305} (\bibinfo {year} {2014})},\
  \Eprint {https://arxiv.org/abs/1309.3794} {arXiv:1309.3794 [nucl-th]}
  \BibitemShut {NoStop}%
\bibitem [{\citenamefont {Weiss}\ \emph
  {et~al.}(2015{\natexlab{a}})\citenamefont {Weiss}, \citenamefont {Bazak},\
  and\ \citenamefont {Barnea}}]{Weiss:2015mba}%
  \BibitemOpen
  \bibfield  {author} {\bibinfo {author} {\bibfnamefont {R.}~\bibnamefont
  {Weiss}}, \bibinfo {author} {\bibfnamefont {B.}~\bibnamefont {Bazak}},\ and\
  \bibinfo {author} {\bibfnamefont {N.}~\bibnamefont {Barnea}},\ }\bibfield
  {title} {\bibinfo {title} {{Generalized nuclear contacts and momentum
  distributions}},\ }\href {https://doi.org/10.1103/PhysRevC.92.054311}
  {\bibfield  {journal} {\bibinfo  {journal} {Phys. Rev.}\ }\textbf {\bibinfo
  {volume} {C92}},\ \bibinfo {pages} {054311} (\bibinfo {year}
  {2015}{\natexlab{a}})},\ \Eprint {https://arxiv.org/abs/1503.07047}
  {arXiv:1503.07047 [nucl-th]} \BibitemShut {NoStop}%
\bibitem [{\citenamefont {Weiss}\ \emph {et~al.}(2018)\citenamefont {Weiss},
  \citenamefont {Cruz-Torres}, \citenamefont {Barnea}, \citenamefont
  {Piasetzky},\ and\ \citenamefont {Hen}}]{Weiss:2016obx}%
  \BibitemOpen
  \bibfield  {author} {\bibinfo {author} {\bibfnamefont {R.}~\bibnamefont
  {Weiss}}, \bibinfo {author} {\bibfnamefont {R.}~\bibnamefont {Cruz-Torres}},
  \bibinfo {author} {\bibfnamefont {N.}~\bibnamefont {Barnea}}, \bibinfo
  {author} {\bibfnamefont {E.}~\bibnamefont {Piasetzky}},\ and\ \bibinfo
  {author} {\bibfnamefont {O.}~\bibnamefont {Hen}},\ }\bibfield  {title}
  {\bibinfo {title} {{The nuclear contacts and short range correlations in
  nuclei}},\ }\href {https://doi.org/10.1016/j.physletb.2018.01.061} {\bibfield
   {journal} {\bibinfo  {journal} {Phys. Lett. B}\ }\textbf {\bibinfo {volume}
  {780}},\ \bibinfo {pages} {211} (\bibinfo {year} {2018})},\ \Eprint
  {https://arxiv.org/abs/1612.00923} {arXiv:1612.00923 [nucl-th]} \BibitemShut
  {NoStop}%
\bibitem [{\citenamefont {Benhar}\ \emph {et~al.}(1994)\citenamefont {Benhar},
  \citenamefont {Fabrocini}, \citenamefont {Fantoni},\ and\ \citenamefont
  {Sick}}]{Benhar:1994hw}%
  \BibitemOpen
  \bibfield  {author} {\bibinfo {author} {\bibfnamefont {O.}~\bibnamefont
  {Benhar}}, \bibinfo {author} {\bibfnamefont {A.}~\bibnamefont {Fabrocini}},
  \bibinfo {author} {\bibfnamefont {S.}~\bibnamefont {Fantoni}},\ and\ \bibinfo
  {author} {\bibfnamefont {I.}~\bibnamefont {Sick}},\ }\bibfield  {title}
  {\bibinfo {title} {{Spectral function of finite nuclei and scattering of GeV
  electrons}},\ }\href {https://doi.org/10.1016/0375-9474(94)90920-2}
  {\bibfield  {journal} {\bibinfo  {journal} {Nucl. Phys. A}\ }\textbf
  {\bibinfo {volume} {579}},\ \bibinfo {pages} {493} (\bibinfo {year}
  {1994})}\BibitemShut {NoStop}%
\bibitem [{\citenamefont {Ciofi~degli Atti}\ and\ \citenamefont
  {Simula}(1996)}]{cda96}%
  \BibitemOpen
  \bibfield  {author} {\bibinfo {author} {\bibfnamefont {C.}~\bibnamefont
  {Ciofi~degli Atti}}\ and\ \bibinfo {author} {\bibfnamefont {S.}~\bibnamefont
  {Simula}},\ }\bibfield  {title} {\bibinfo {title} {Realistic model of the
  nucleon spectral function in few- and many-nucleon systems},\ }\href
  {https://doi.org/10.1103/PhysRevC.53.1689} {\bibfield  {journal} {\bibinfo
  {journal} {Phys. Rev. C}\ }\textbf {\bibinfo {volume} {53}},\ \bibinfo
  {pages} {1689} (\bibinfo {year} {1996})}\BibitemShut {NoStop}%
\bibitem [{\citenamefont {Weiss}\ \emph {et~al.}(2019)\citenamefont {Weiss},
  \citenamefont {Korover}, \citenamefont {Piasetzky}, \citenamefont {Hen},\
  and\ \citenamefont {Barnea}}]{Weiss:2018tbu}%
  \BibitemOpen
  \bibfield  {author} {\bibinfo {author} {\bibfnamefont {R.}~\bibnamefont
  {Weiss}}, \bibinfo {author} {\bibfnamefont {I.}~\bibnamefont {Korover}},
  \bibinfo {author} {\bibfnamefont {E.}~\bibnamefont {Piasetzky}}, \bibinfo
  {author} {\bibfnamefont {O.}~\bibnamefont {Hen}},\ and\ \bibinfo {author}
  {\bibfnamefont {N.}~\bibnamefont {Barnea}},\ }\bibfield  {title} {\bibinfo
  {title} {{Energy and momentum dependence of nuclear short-range correlations
  - Spectral function, exclusive scattering experiments and the contact
  formalism}},\ }\href {https://doi.org/10.1016/j.physletb.2019.02.019}
  {\bibfield  {journal} {\bibinfo  {journal} {Phys. Lett.}\ }\textbf {\bibinfo
  {volume} {B791}},\ \bibinfo {pages} {242} (\bibinfo {year} {2019})},\ \Eprint
  {https://arxiv.org/abs/1806.10217} {arXiv:1806.10217 [nucl-th]} \BibitemShut
  {NoStop}%
\bibitem [{\citenamefont {Pastore}\ \emph {et~al.}(2020)\citenamefont
  {Pastore}, \citenamefont {Carlson}, \citenamefont {Gandolfi}, \citenamefont
  {Schiavilla},\ and\ \citenamefont {Wiringa}}]{Pastore:2019urn}%
  \BibitemOpen
  \bibfield  {author} {\bibinfo {author} {\bibfnamefont {S.}~\bibnamefont
  {Pastore}}, \bibinfo {author} {\bibfnamefont {J.}~\bibnamefont {Carlson}},
  \bibinfo {author} {\bibfnamefont {S.}~\bibnamefont {Gandolfi}}, \bibinfo
  {author} {\bibfnamefont {R.}~\bibnamefont {Schiavilla}},\ and\ \bibinfo
  {author} {\bibfnamefont {R.~B.}\ \bibnamefont {Wiringa}},\ }\bibfield
  {title} {\bibinfo {title} {{Quasielastic lepton scattering and back-to-back
  nucleons in the short-time approximation}},\ }\href
  {https://doi.org/10.1103/PhysRevC.101.044612} {\bibfield  {journal} {\bibinfo
   {journal} {Phys. Rev. C}\ }\textbf {\bibinfo {volume} {101}},\ \bibinfo
  {pages} {044612} (\bibinfo {year} {2020})},\ \Eprint
  {https://arxiv.org/abs/1909.06400} {arXiv:1909.06400 [nucl-th]} \BibitemShut
  {NoStop}%
\bibitem [{\citenamefont {Andreoli}\ \emph {et~al.}(2022)\citenamefont
  {Andreoli}, \citenamefont {Carlson}, \citenamefont {Lovato}, \citenamefont
  {Pastore}, \citenamefont {Rocco},\ and\ \citenamefont
  {Wiringa}}]{Andreoli:2021cxo}%
  \BibitemOpen
  \bibfield  {author} {\bibinfo {author} {\bibfnamefont {L.}~\bibnamefont
  {Andreoli}}, \bibinfo {author} {\bibfnamefont {J.}~\bibnamefont {Carlson}},
  \bibinfo {author} {\bibfnamefont {A.}~\bibnamefont {Lovato}}, \bibinfo
  {author} {\bibfnamefont {S.}~\bibnamefont {Pastore}}, \bibinfo {author}
  {\bibfnamefont {N.}~\bibnamefont {Rocco}},\ and\ \bibinfo {author}
  {\bibfnamefont {R.~B.}\ \bibnamefont {Wiringa}},\ }\bibfield  {title}
  {\bibinfo {title} {{Electron scattering on A=3 nuclei from quantum Monte
  Carlo based approaches}},\ }\href
  {https://doi.org/10.1103/PhysRevC.105.014002} {\bibfield  {journal} {\bibinfo
   {journal} {Phys. Rev. C}\ }\textbf {\bibinfo {volume} {105}},\ \bibinfo
  {pages} {014002} (\bibinfo {year} {2022})},\ \Eprint
  {https://arxiv.org/abs/2108.10824} {arXiv:2108.10824 [nucl-th]} \BibitemShut
  {NoStop}%
\bibitem [{\citenamefont {Pandharipande}\ \emph {et~al.}(1997)\citenamefont
  {Pandharipande}, \citenamefont {Sick},\ and\ \citenamefont
  {Huberts}}]{Pandharipande:1997zz}%
  \BibitemOpen
  \bibfield  {author} {\bibinfo {author} {\bibfnamefont {V.~R.}\ \bibnamefont
  {Pandharipande}}, \bibinfo {author} {\bibfnamefont {I.}~\bibnamefont
  {Sick}},\ and\ \bibinfo {author} {\bibfnamefont {P.~K. A.~d.}\ \bibnamefont
  {Huberts}},\ }\bibfield  {title} {\bibinfo {title} {{Independent particle
  motion and correlations in fermion systems}},\ }\href
  {https://doi.org/10.1103/RevModPhys.69.981} {\bibfield  {journal} {\bibinfo
  {journal} {Rev. Mod. Phys.}\ }\textbf {\bibinfo {volume} {69}},\ \bibinfo
  {pages} {981} (\bibinfo {year} {1997})}\BibitemShut {NoStop}%
\bibitem [{\citenamefont {Dickhoff}\ and\ \citenamefont
  {Barbieri}(2004)}]{Dickhoff:2004xx}%
  \BibitemOpen
  \bibfield  {author} {\bibinfo {author} {\bibfnamefont {W.~H.}\ \bibnamefont
  {Dickhoff}}\ and\ \bibinfo {author} {\bibfnamefont {C.}~\bibnamefont
  {Barbieri}},\ }\bibfield  {title} {\bibinfo {title} {{Selfconsistent Green's
  function method for nuclei and nuclear matter}},\ }\href
  {https://doi.org/10.1016/j.ppnp.2004.02.038} {\bibfield  {journal} {\bibinfo
  {journal} {Prog. Part. Nucl. Phys.}\ }\textbf {\bibinfo {volume} {52}},\
  \bibinfo {pages} {377} (\bibinfo {year} {2004})},\ \Eprint
  {https://arxiv.org/abs/nucl-th/0402034} {arXiv:nucl-th/0402034} \BibitemShut
  {NoStop}%
\bibitem [{\citenamefont {Lapikas}(1993)}]{Lapikas:1993uwd}%
  \BibitemOpen
  \bibfield  {author} {\bibinfo {author} {\bibfnamefont {L.}~\bibnamefont
  {Lapikas}},\ }\bibfield  {title} {\bibinfo {title} {{Quasi-elastic electron
  scattering off nuclei}},\ }\href
  {https://doi.org/10.1016/0375-9474(93)90630-G} {\bibfield  {journal}
  {\bibinfo  {journal} {Nucl. Phys. A}\ }\textbf {\bibinfo {volume} {553}},\
  \bibinfo {pages} {297c} (\bibinfo {year} {1993})}\BibitemShut {NoStop}%
\bibitem [{\citenamefont {Kramer}\ \emph {et~al.}(2001)\citenamefont {Kramer},
  \citenamefont {Blok},\ and\ \citenamefont {Lapikas}}]{Kramer:2000kc}%
  \BibitemOpen
  \bibfield  {author} {\bibinfo {author} {\bibfnamefont {G.~J.}\ \bibnamefont
  {Kramer}}, \bibinfo {author} {\bibfnamefont {H.~P.}\ \bibnamefont {Blok}},\
  and\ \bibinfo {author} {\bibfnamefont {L.}~\bibnamefont {Lapikas}},\
  }\bibfield  {title} {\bibinfo {title} {{A Consistent analysis of (e,e-prime
  p) and (d,He-3) experiments}},\ }\href
  {https://doi.org/10.1016/S0375-9474(00)00379-1} {\bibfield  {journal}
  {\bibinfo  {journal} {Nucl. Phys. A}\ }\textbf {\bibinfo {volume} {679}},\
  \bibinfo {pages} {267} (\bibinfo {year} {2001})},\ \Eprint
  {https://arxiv.org/abs/nucl-ex/0007014} {arXiv:nucl-ex/0007014} \BibitemShut
  {NoStop}%
\bibitem [{\citenamefont {Gao}\ \emph {et~al.}(2000)\citenamefont {Gao} \emph
  {et~al.}}]{JeffersonLabHallA:2000dxx}%
  \BibitemOpen
  \bibfield  {author} {\bibinfo {author} {\bibfnamefont {J.}~\bibnamefont
  {Gao}} \emph {et~al.} (\bibinfo {collaboration} {Jefferson Lab Hall A}),\
  }\bibfield  {title} {\bibinfo {title} {{Dynamical relativistic effects in
  quasielastic 1p shell knockout}},\ }\href
  {https://doi.org/10.1103/PhysRevLett.84.3265} {\bibfield  {journal} {\bibinfo
   {journal} {Phys. Rev. Lett.}\ }\textbf {\bibinfo {volume} {84}},\ \bibinfo
  {pages} {3265} (\bibinfo {year} {2000})}\BibitemShut {NoStop}%
\bibitem [{\citenamefont {Gade}\ \emph {et~al.}(2008)\citenamefont {Gade} \emph
  {et~al.}}]{Gade:2008zz}%
  \BibitemOpen
  \bibfield  {author} {\bibinfo {author} {\bibfnamefont {A.}~\bibnamefont
  {Gade}} \emph {et~al.},\ }\bibfield  {title} {\bibinfo {title} {{Reduction of
  spectroscopic strength: Weakly-bound and strongly-bound single-particle
  states studied using one-nucleon knockout reactions}},\ }\href
  {https://doi.org/10.1103/PhysRevC.77.044306} {\bibfield  {journal} {\bibinfo
  {journal} {Phys. Rev. C}\ }\textbf {\bibinfo {volume} {77}},\ \bibinfo
  {pages} {044306} (\bibinfo {year} {2008})}\BibitemShut {NoStop}%
\bibitem [{\citenamefont {Flavigny}\ \emph {et~al.}(2013)\citenamefont
  {Flavigny} \emph {et~al.}}]{Flavigny:2013bha}%
  \BibitemOpen
  \bibfield  {author} {\bibinfo {author} {\bibfnamefont {F.}~\bibnamefont
  {Flavigny}} \emph {et~al.},\ }\bibfield  {title} {\bibinfo {title} {{Limited
  Asymmetry Dependence of Correlations from Single Nucleon Transfer}},\ }\href
  {https://doi.org/10.1103/PhysRevLett.110.122503} {\bibfield  {journal}
  {\bibinfo  {journal} {Phys. Rev. Lett.}\ }\textbf {\bibinfo {volume} {110}},\
  \bibinfo {pages} {122503} (\bibinfo {year} {2013})}\BibitemShut {NoStop}%
\bibitem [{\citenamefont {Kay}\ \emph {et~al.}(2013)\citenamefont {Kay},
  \citenamefont {Schiffer},\ and\ \citenamefont {Freeman}}]{Kay:2013bsa}%
  \BibitemOpen
  \bibfield  {author} {\bibinfo {author} {\bibfnamefont {B.~P.}\ \bibnamefont
  {Kay}}, \bibinfo {author} {\bibfnamefont {J.~P.}\ \bibnamefont {Schiffer}},\
  and\ \bibinfo {author} {\bibfnamefont {S.~J.}\ \bibnamefont {Freeman}},\
  }\bibfield  {title} {\bibinfo {title} {{Quenching of Cross Sections in
  Nucleon Transfer Reactions}},\ }\href
  {https://doi.org/10.1103/PhysRevLett.111.042502} {\bibfield  {journal}
  {\bibinfo  {journal} {Phys. Rev. Lett.}\ }\textbf {\bibinfo {volume} {111}},\
  \bibinfo {pages} {042502} (\bibinfo {year} {2013})},\ \Eprint
  {https://arxiv.org/abs/1307.1178} {arXiv:1307.1178 [nucl-ex]} \BibitemShut
  {NoStop}%
\bibitem [{\citenamefont {Tostevin}\ and\ \citenamefont
  {Gade}(2014)}]{Tostevin:2014usa}%
  \BibitemOpen
  \bibfield  {author} {\bibinfo {author} {\bibfnamefont {J.~A.}\ \bibnamefont
  {Tostevin}}\ and\ \bibinfo {author} {\bibfnamefont {A.}~\bibnamefont
  {Gade}},\ }\bibfield  {title} {\bibinfo {title} {{Systematics of
  intermediate-energy single-nucleon removal cross sections}},\ }\href
  {https://doi.org/10.1103/PhysRevC.90.057602} {\bibfield  {journal} {\bibinfo
  {journal} {Phys. Rev. C}\ }\textbf {\bibinfo {volume} {90}},\ \bibinfo
  {pages} {057602} (\bibinfo {year} {2014})},\ \Eprint
  {https://arxiv.org/abs/1409.6576} {arXiv:1409.6576 [nucl-th]} \BibitemShut
  {NoStop}%
\bibitem [{\citenamefont {Atar}\ \emph {et~al.}(2018)\citenamefont {Atar} \emph
  {et~al.}}]{Atar:2018dhg}%
  \BibitemOpen
  \bibfield  {author} {\bibinfo {author} {\bibfnamefont {L.}~\bibnamefont
  {Atar}} \emph {et~al.},\ }\bibfield  {title} {\bibinfo {title} {{Quasifree
  ($p$, $2p$) Reactions on Oxygen Isotopes: Observation of Isospin Independence
  of the Reduced Single-Particle Strength}},\ }\href
  {https://doi.org/10.1103/PhysRevLett.120.052501} {\bibfield  {journal}
  {\bibinfo  {journal} {Phys. Rev. Lett.}\ }\textbf {\bibinfo {volume} {120}},\
  \bibinfo {pages} {052501} (\bibinfo {year} {2018})}\BibitemShut {NoStop}%
\bibitem [{\citenamefont {Kawase}\ \emph {et~al.}(2018)\citenamefont {Kawase}
  \emph {et~al.}}]{Kawase:2018ojr}%
  \BibitemOpen
  \bibfield  {author} {\bibinfo {author} {\bibfnamefont {S.}~\bibnamefont
  {Kawase}} \emph {et~al.},\ }\bibfield  {title} {\bibinfo {title} {{Exclusive
  quasi-free proton knockout from oxygen isotopes at intermediate energies}},\
  }\href {https://doi.org/10.1093/ptep/pty011} {\bibfield  {journal} {\bibinfo
  {journal} {PTEP}\ }\textbf {\bibinfo {volume} {2018}},\ \bibinfo {pages}
  {021D01} (\bibinfo {year} {2018})}\BibitemShut {NoStop}%
\bibitem [{\citenamefont {Paschalis}\ \emph {et~al.}(2020)\citenamefont
  {Paschalis}, \citenamefont {Petri}, \citenamefont {Macchiavelli},
  \citenamefont {Hen},\ and\ \citenamefont {Piasetzky}}]{Paschalis:2018zkx}%
  \BibitemOpen
  \bibfield  {author} {\bibinfo {author} {\bibfnamefont {S.}~\bibnamefont
  {Paschalis}}, \bibinfo {author} {\bibfnamefont {M.}~\bibnamefont {Petri}},
  \bibinfo {author} {\bibfnamefont {A.~O.}\ \bibnamefont {Macchiavelli}},
  \bibinfo {author} {\bibfnamefont {O.}~\bibnamefont {Hen}},\ and\ \bibinfo
  {author} {\bibfnamefont {E.}~\bibnamefont {Piasetzky}},\ }\bibfield  {title}
  {\bibinfo {title} {{Nucleon-nucleon correlations and the single-particle
  strength in atomic nuclei}},\ }\href
  {https://doi.org/10.1016/j.physletb.2019.135110} {\bibfield  {journal}
  {\bibinfo  {journal} {Phys. Lett. B}\ }\textbf {\bibinfo {volume} {800}},\
  \bibinfo {pages} {135110} (\bibinfo {year} {2020})},\ \Eprint
  {https://arxiv.org/abs/1812.08051} {arXiv:1812.08051 [nucl-ex]} \BibitemShut
  {NoStop}%
\bibitem [{\citenamefont {Atkinson}\ \emph {et~al.}(2018)\citenamefont
  {Atkinson}, \citenamefont {Blok}, \citenamefont {Lapik\'as}, \citenamefont
  {Charity},\ and\ \citenamefont {Dickhoff}}]{Atkinson:2018nvp}%
  \BibitemOpen
  \bibfield  {author} {\bibinfo {author} {\bibfnamefont {M.~C.}\ \bibnamefont
  {Atkinson}}, \bibinfo {author} {\bibfnamefont {H.~P.}\ \bibnamefont {Blok}},
  \bibinfo {author} {\bibfnamefont {L.}~\bibnamefont {Lapik\'as}}, \bibinfo
  {author} {\bibfnamefont {R.~J.}\ \bibnamefont {Charity}},\ and\ \bibinfo
  {author} {\bibfnamefont {W.~H.}\ \bibnamefont {Dickhoff}},\ }\bibfield
  {title} {\bibinfo {title} {{Validity of the distorted-wave
  impulse-approximation description of $^{40}Ca(e,e'p)^{39}K$ data using only
  ingredients from a nonlocal dispersive optical model}},\ }\href
  {https://doi.org/10.1103/PhysRevC.98.044627} {\bibfield  {journal} {\bibinfo
  {journal} {Phys. Rev. C}\ }\textbf {\bibinfo {volume} {98}},\ \bibinfo
  {pages} {044627} (\bibinfo {year} {2018})},\ \Eprint
  {https://arxiv.org/abs/1808.08895} {arXiv:1808.08895 [nucl-th]} \BibitemShut
  {NoStop}%
\bibitem [{\citenamefont {Simkovic}\ \emph {et~al.}(2009)\citenamefont
  {Simkovic}, \citenamefont {Faessler}, \citenamefont {Muther}, \citenamefont
  {Rodin},\ and\ \citenamefont {Stauf}}]{Simkovic:2009pp}%
  \BibitemOpen
  \bibfield  {author} {\bibinfo {author} {\bibfnamefont {F.}~\bibnamefont
  {Simkovic}}, \bibinfo {author} {\bibfnamefont {A.}~\bibnamefont {Faessler}},
  \bibinfo {author} {\bibfnamefont {H.}~\bibnamefont {Muther}}, \bibinfo
  {author} {\bibfnamefont {V.}~\bibnamefont {Rodin}},\ and\ \bibinfo {author}
  {\bibfnamefont {M.}~\bibnamefont {Stauf}},\ }\bibfield  {title} {\bibinfo
  {title} {{The 0 nu bb-decay nuclear matrix elements with self-consistent
  short-range correlations}},\ }\href
  {https://doi.org/10.1103/PhysRevC.79.055501} {\bibfield  {journal} {\bibinfo
  {journal} {Phys. Rev.}\ }\textbf {\bibinfo {volume} {C79}},\ \bibinfo {pages}
  {055501} (\bibinfo {year} {2009})},\ \Eprint
  {https://arxiv.org/abs/0902.0331} {arXiv:0902.0331 [nucl-th]} \BibitemShut
  {NoStop}%
\bibitem [{\citenamefont {Cirigliano}\ \emph {et~al.}(2018)\citenamefont
  {Cirigliano}, \citenamefont {Dekens}, \citenamefont {de~Vries}, \citenamefont
  {Graesser}, \citenamefont {Mereghetti}, \citenamefont {Pastore},\ and\
  \citenamefont {van Kolck}}]{PhysRevLett.120.202001}%
  \BibitemOpen
  \bibfield  {author} {\bibinfo {author} {\bibfnamefont {V.}~\bibnamefont
  {Cirigliano}}, \bibinfo {author} {\bibfnamefont {W.}~\bibnamefont {Dekens}},
  \bibinfo {author} {\bibfnamefont {J.}~\bibnamefont {de~Vries}}, \bibinfo
  {author} {\bibfnamefont {M.~L.}\ \bibnamefont {Graesser}}, \bibinfo {author}
  {\bibfnamefont {E.}~\bibnamefont {Mereghetti}}, \bibinfo {author}
  {\bibfnamefont {S.}~\bibnamefont {Pastore}},\ and\ \bibinfo {author}
  {\bibfnamefont {U.}~\bibnamefont {van Kolck}},\ }\bibfield  {title} {\bibinfo
  {title} {New leading contribution to neutrinoless double-$\ensuremath{\beta}$
  decay},\ }\href {https://doi.org/10.1103/PhysRevLett.120.202001} {\bibfield
  {journal} {\bibinfo  {journal} {Phys. Rev. Lett.}\ }\textbf {\bibinfo
  {volume} {120}},\ \bibinfo {pages} {202001} (\bibinfo {year}
  {2018})}\BibitemShut {NoStop}%
\bibitem [{\citenamefont {Weiss}\ \emph {et~al.}(2022)\citenamefont {Weiss},
  \citenamefont {Soriano}, \citenamefont {Lovato}, \citenamefont {Menendez},\
  and\ \citenamefont {Wiringa}}]{Weiss:2021rig}%
  \BibitemOpen
  \bibfield  {author} {\bibinfo {author} {\bibfnamefont {R.}~\bibnamefont
  {Weiss}}, \bibinfo {author} {\bibfnamefont {P.}~\bibnamefont {Soriano}},
  \bibinfo {author} {\bibfnamefont {A.}~\bibnamefont {Lovato}}, \bibinfo
  {author} {\bibfnamefont {J.}~\bibnamefont {Menendez}},\ and\ \bibinfo
  {author} {\bibfnamefont {R.~B.}\ \bibnamefont {Wiringa}},\ }\bibfield
  {title} {\bibinfo {title} {{Neutrinoless double-\ensuremath{\beta} decay:
  Combining quantum Monte Carlo and the nuclear shell model with the
  generalized contact formalism}},\ }\href
  {https://doi.org/10.1103/PhysRevC.106.065501} {\bibfield  {journal} {\bibinfo
   {journal} {Phys. Rev. C}\ }\textbf {\bibinfo {volume} {106}},\ \bibinfo
  {pages} {065501} (\bibinfo {year} {2022})},\ \Eprint
  {https://arxiv.org/abs/2112.08146} {arXiv:2112.08146 [nucl-th]} \BibitemShut
  {NoStop}%
\bibitem [{\citenamefont {Hen}\ \emph {et~al.}(2015)\citenamefont {Hen},
  \citenamefont {Li}, \citenamefont {Guo}, \citenamefont {Weinstein},\ and\
  \citenamefont {Piasetzky}}]{hen15}%
  \BibitemOpen
  \bibfield  {author} {\bibinfo {author} {\bibfnamefont {O.}~\bibnamefont
  {Hen}}, \bibinfo {author} {\bibfnamefont {B.-A.}\ \bibnamefont {Li}},
  \bibinfo {author} {\bibfnamefont {W.-J.}\ \bibnamefont {Guo}}, \bibinfo
  {author} {\bibfnamefont {L.~B.}\ \bibnamefont {Weinstein}},\ and\ \bibinfo
  {author} {\bibfnamefont {E.}~\bibnamefont {Piasetzky}},\ }\bibfield  {title}
  {\bibinfo {title} {Symmetry energy of nucleonic matter with tensor
  correlations},\ }\href {https://doi.org/10.1103/PhysRevC.91.025803}
  {\bibfield  {journal} {\bibinfo  {journal} {Phys. Rev. C}\ }\textbf {\bibinfo
  {volume} {91}},\ \bibinfo {pages} {025803} (\bibinfo {year}
  {2015})}\BibitemShut {NoStop}%
\bibitem [{\citenamefont {Cai}\ and\ \citenamefont {Li}(2016)}]{Cai:2015xga}%
  \BibitemOpen
  \bibfield  {author} {\bibinfo {author} {\bibfnamefont {B.-J.}\ \bibnamefont
  {Cai}}\ and\ \bibinfo {author} {\bibfnamefont {B.-A.}\ \bibnamefont {Li}},\
  }\bibfield  {title} {\bibinfo {title} {{Symmetry energy of cold nucleonic
  matter within a relativistic mean field model encapsulating effects of high
  momentum nucleons induced by short-range correlations}},\ }\href
  {https://doi.org/10.1103/PhysRevC.93.014619} {\bibfield  {journal} {\bibinfo
  {journal} {Phys. Rev.}\ }\textbf {\bibinfo {volume} {C93}},\ \bibinfo {pages}
  {014619} (\bibinfo {year} {2016})}\BibitemShut {NoStop}%
\bibitem [{\citenamefont {Li}\ \emph {et~al.}(2018)\citenamefont {Li},
  \citenamefont {Cai}, \citenamefont {Chen},\ and\ \citenamefont
  {Xu}}]{Li:2018lpy}%
  \BibitemOpen
  \bibfield  {author} {\bibinfo {author} {\bibfnamefont {B.-A.}\ \bibnamefont
  {Li}}, \bibinfo {author} {\bibfnamefont {B.-J.}\ \bibnamefont {Cai}},
  \bibinfo {author} {\bibfnamefont {L.-W.}\ \bibnamefont {Chen}},\ and\
  \bibinfo {author} {\bibfnamefont {J.}~\bibnamefont {Xu}},\ }\bibfield
  {title} {\bibinfo {title} {{Nucleon Effective Masses in Neutron-Rich
  Matter}},\ }\href {https://doi.org/10.1016/j.ppnp.2018.01.001} {\bibfield
  {journal} {\bibinfo  {journal} {Prog. Part. Nucl. Phys.}\ }\textbf {\bibinfo
  {volume} {99}},\ \bibinfo {pages} {29} (\bibinfo {year} {2018})},\ \Eprint
  {https://arxiv.org/abs/1801.01213} {arXiv:1801.01213 [nucl-th]} \BibitemShut
  {NoStop}%
\bibitem [{\citenamefont {Carbone}\ \emph {et~al.}(2012)\citenamefont
  {Carbone}, \citenamefont {Polls},\ and\ \citenamefont {Rios}}]{Carbone_2012}%
  \BibitemOpen
  \bibfield  {author} {\bibinfo {author} {\bibfnamefont {A.}~\bibnamefont
  {Carbone}}, \bibinfo {author} {\bibfnamefont {A.}~\bibnamefont {Polls}},\
  and\ \bibinfo {author} {\bibfnamefont {A.}~\bibnamefont {Rios}},\ }\bibfield
  {title} {\bibinfo {title} {High-momentum components in the nuclear symmetry
  energy},\ }\href {https://doi.org/10.1209/0295-5075/97/22001} {\bibfield
  {journal} {\bibinfo  {journal} {Europhysics Letters}\ }\textbf {\bibinfo
  {volume} {97}},\ \bibinfo {pages} {22001} (\bibinfo {year}
  {2012})}\BibitemShut {NoStop}%
\bibitem [{\citenamefont {Ciofi~degli Atti}(2015)}]{Atti:2015eda}%
  \BibitemOpen
  \bibfield  {author} {\bibinfo {author} {\bibfnamefont {C.}~\bibnamefont
  {Ciofi~degli Atti}},\ }\bibfield  {title} {\bibinfo {title} {{In-medium
  short-range dynamics of nucleons: Recent theoretical and experimental
  advances}},\ }\href {https://doi.org/10.1016/j.physrep.2015.06.002}
  {\bibfield  {journal} {\bibinfo  {journal} {Phys. Rept.}\ }\textbf {\bibinfo
  {volume} {590}},\ \bibinfo {pages} {1} (\bibinfo {year} {2015})}\BibitemShut
  {NoStop}%
\bibitem [{\citenamefont {Hen}\ \emph {et~al.}(2017)\citenamefont {Hen},
  \citenamefont {Miller}, \citenamefont {Piasetzky},\ and\ \citenamefont
  {Weinstein}}]{Hen:2016kwk}%
  \BibitemOpen
  \bibfield  {author} {\bibinfo {author} {\bibfnamefont {O.}~\bibnamefont
  {Hen}}, \bibinfo {author} {\bibfnamefont {G.~A.}\ \bibnamefont {Miller}},
  \bibinfo {author} {\bibfnamefont {E.}~\bibnamefont {Piasetzky}},\ and\
  \bibinfo {author} {\bibfnamefont {L.~B.}\ \bibnamefont {Weinstein}},\
  }\bibfield  {title} {\bibinfo {title} {{Nucleon-Nucleon Correlations,
  Short-lived Excitations, and the Quarks Within}},\ }\href
  {https://doi.org/10.1103/RevModPhys.89.045002} {\bibfield  {journal}
  {\bibinfo  {journal} {Rev. Mod. Phys.}\ }\textbf {\bibinfo {volume} {89}},\
  \bibinfo {pages} {045002} (\bibinfo {year} {2017})}\BibitemShut {NoStop}%
\bibitem [{\citenamefont {Arrington}\ \emph {et~al.}(2022)\citenamefont
  {Arrington}, \citenamefont {Fomin},\ and\ \citenamefont
  {Schmidt}}]{Arrington:2022sov}%
  \BibitemOpen
  \bibfield  {author} {\bibinfo {author} {\bibfnamefont {J.}~\bibnamefont
  {Arrington}}, \bibinfo {author} {\bibfnamefont {N.}~\bibnamefont {Fomin}},\
  and\ \bibinfo {author} {\bibfnamefont {A.}~\bibnamefont {Schmidt}},\
  }\bibfield  {title} {\bibinfo {title} {Progress in understanding short-range
  structure in nuclei: An experimental perspective},\ }\href
  {https://doi.org/10.1146/annurev-nucl-102020-022253} {\bibfield  {journal}
  {\bibinfo  {journal} {Annual Review of Nuclear and Particle Science}\
  }\textbf {\bibinfo {volume} {72}},\ \bibinfo {pages} {307} (\bibinfo {year}
  {2022})}\BibitemShut {NoStop}%
\bibitem [{\citenamefont {Frankfurt}\ and\ \citenamefont
  {Strikman}(1981)}]{Frankfurt81}%
  \BibitemOpen
  \bibfield  {author} {\bibinfo {author} {\bibfnamefont {L.~L.}\ \bibnamefont
  {Frankfurt}}\ and\ \bibinfo {author} {\bibfnamefont {M.~I.}\ \bibnamefont
  {Strikman}},\ }\bibfield  {title} {\bibinfo {title} {High-energy phenomena,
  short-range nuclear structure and qcd},\ }\href@noop {} {\bibfield  {journal}
  {\bibinfo  {journal} {Phys. Rep.}\ }\textbf {\bibinfo {volume} {76}},\
  \bibinfo {pages} {215} (\bibinfo {year} {1981})}\BibitemShut {NoStop}%
\bibitem [{\citenamefont {Frankfurt}\ and\ \citenamefont
  {Strikman}(1988)}]{Frankfurt88}%
  \BibitemOpen
  \bibfield  {author} {\bibinfo {author} {\bibfnamefont {L.}~\bibnamefont
  {Frankfurt}}\ and\ \bibinfo {author} {\bibfnamefont {M.}~\bibnamefont
  {Strikman}},\ }\bibfield  {title} {\bibinfo {title} {Hard nuclear processes
  and microscopic nuclear structure},\ }\href@noop {} {\bibfield  {journal}
  {\bibinfo  {journal} {Phys. Rep.}\ }\textbf {\bibinfo {volume} {160}},\
  \bibinfo {pages} {235 } (\bibinfo {year} {1988})}\BibitemShut {NoStop}%
\bibitem [{\citenamefont {Frankfurt}\ \emph {et~al.}(1993)\citenamefont
  {Frankfurt}, \citenamefont {Strikman}, \citenamefont {Day},\ and\
  \citenamefont {Sargsyan}}]{frankfurt93}%
  \BibitemOpen
  \bibfield  {author} {\bibinfo {author} {\bibfnamefont {L.}~\bibnamefont
  {Frankfurt}}, \bibinfo {author} {\bibfnamefont {M.}~\bibnamefont {Strikman}},
  \bibinfo {author} {\bibfnamefont {D.}~\bibnamefont {Day}},\ and\ \bibinfo
  {author} {\bibfnamefont {M.}~\bibnamefont {Sargsyan}},\ }\bibfield  {title}
  {\bibinfo {title} {Evidence for short-range correlations from high q2 (e,e')
  reactions},\ }\href@noop {} {\bibfield  {journal} {\bibinfo  {journal} {Phys.
  Rev. C}\ }\textbf {\bibinfo {volume} {48}},\ \bibinfo {pages} {2451}
  (\bibinfo {year} {1993})}\BibitemShut {NoStop}%
\bibitem [{\citenamefont {Egiyan}\ \emph {et~al.}(2003)\citenamefont {Egiyan}
  \emph {et~al.}}]{egiyan02}%
  \BibitemOpen
  \bibfield  {author} {\bibinfo {author} {\bibfnamefont {K.}~\bibnamefont
  {Egiyan}} \emph {et~al.} (\bibinfo {collaboration} {CLAS Collaboration}),\
  }\href@noop {} {\bibfield  {journal} {\bibinfo  {journal} {Phys. Rev. C}\
  }\textbf {\bibinfo {volume} {68}},\ \bibinfo {pages} {014313} (\bibinfo
  {year} {2003})}\BibitemShut {NoStop}%
\bibitem [{\citenamefont {Egiyan}\ \emph {et~al.}(2006)\citenamefont {Egiyan}
  \emph {et~al.}}]{egiyan06}%
  \BibitemOpen
  \bibfield  {author} {\bibinfo {author} {\bibfnamefont {K.}~\bibnamefont
  {Egiyan}} \emph {et~al.} (\bibinfo {collaboration} {CLAS Collaboration}),\
  }\bibfield  {title} {\bibinfo {title} {Measurement of 2- and 3-nucleon short
  range correlation probabilities in nuclei},\ }\href@noop {} {\bibfield
  {journal} {\bibinfo  {journal} {Phys. Rev. Lett.}\ }\textbf {\bibinfo
  {volume} {96}},\ \bibinfo {pages} {082501} (\bibinfo {year}
  {2006})}\BibitemShut {NoStop}%
\bibitem [{\citenamefont {Fomin}\ \emph {et~al.}(2012)\citenamefont {Fomin}
  \emph {et~al.}}]{fomin12}%
  \BibitemOpen
  \bibfield  {author} {\bibinfo {author} {\bibfnamefont {N.}~\bibnamefont
  {Fomin}} \emph {et~al.},\ }\bibfield  {title} {\bibinfo {title} {New
  measurements of high-momentum nucleons and short-range structures in
  nuclei},\ }\href@noop {} {\bibfield  {journal} {\bibinfo  {journal} {Phys.
  Rev. Lett.}\ }\textbf {\bibinfo {volume} {108}},\ \bibinfo {pages} {092502}
  (\bibinfo {year} {2012})}\BibitemShut {NoStop}%
\bibitem [{\citenamefont {Schmookler}\ \emph {et~al.}(2019)\citenamefont
  {Schmookler} \emph {et~al.}}]{Schmookler:2019nvf}%
  \BibitemOpen
  \bibfield  {author} {\bibinfo {author} {\bibfnamefont {B.}~\bibnamefont
  {Schmookler}} \emph {et~al.} (\bibinfo {collaboration} {CLAS
  Collaboration}),\ }\bibfield  {title} {\bibinfo {title} {{Modified structure
  of protons and neutrons in correlated pairs}},\ }\href
  {https://doi.org/10.1038/s41586-019-0925-9} {\bibfield  {journal} {\bibinfo
  {journal} {Nature}\ }\textbf {\bibinfo {volume} {566}},\ \bibinfo {pages}
  {354} (\bibinfo {year} {2019})}\BibitemShut {NoStop}%
\bibitem [{\citenamefont {Nguyen}\ \emph {et~al.}(2020)\citenamefont {Nguyen}
  \emph {et~al.}}]{JeffersonLabHallA:2020wrr}%
  \BibitemOpen
  \bibfield  {author} {\bibinfo {author} {\bibfnamefont {D.}~\bibnamefont
  {Nguyen}} \emph {et~al.} (\bibinfo {collaboration} {Jefferson Lab Hall A}),\
  }\bibfield  {title} {\bibinfo {title} {{Novel observation of isospin
  structure of short-range correlations in calcium isotopes}},\ }\href
  {https://doi.org/10.1103/PhysRevC.102.064004} {\bibfield  {journal} {\bibinfo
   {journal} {Phys. Rev. C}\ }\textbf {\bibinfo {volume} {102}},\ \bibinfo
  {pages} {064004} (\bibinfo {year} {2020})},\ \Eprint
  {https://arxiv.org/abs/2004.11448} {arXiv:2004.11448 [nucl-ex]} \BibitemShut
  {NoStop}%
\bibitem [{\citenamefont {Li}\ \emph {et~al.}(2022)\citenamefont {Li} \emph
  {et~al.}}]{Li:2022fhh}%
  \BibitemOpen
  \bibfield  {author} {\bibinfo {author} {\bibfnamefont {S.}~\bibnamefont {Li}}
  \emph {et~al.},\ }\bibfield  {title} {\bibinfo {title} {{Revealing the
  short-range structure of the mirror nuclei $^{3}$H and $^{3}$He}},\ }\href
  {https://doi.org/10.1038/s41586-022-05007-2} {\bibfield  {journal} {\bibinfo
  {journal} {Nature}\ }\textbf {\bibinfo {volume} {609}},\ \bibinfo {pages}
  {41} (\bibinfo {year} {2022})},\ \Eprint {https://arxiv.org/abs/2210.04189}
  {arXiv:2210.04189 [nucl-ex]} \BibitemShut {NoStop}%
\bibitem [{\citenamefont {Patsyuk}\ \emph {et~al.}(2021)\citenamefont
  {Patsyuk}, \citenamefont {Kahlbow}, \citenamefont {Laskaris}, \citenamefont
  {Duer}, \citenamefont {Lenivenko}, \citenamefont {Segarra}, \citenamefont
  {Atovullaev}, \citenamefont {Johansson}, \citenamefont {Aumann},
  \citenamefont {Corsi},\ and\ \citenamefont
  {et~al.}}]{patsyuk2021unperturbed}%
  \BibitemOpen
  \bibfield  {author} {\bibinfo {author} {\bibfnamefont {M.}~\bibnamefont
  {Patsyuk}}, \bibinfo {author} {\bibfnamefont {J.}~\bibnamefont {Kahlbow}},
  \bibinfo {author} {\bibfnamefont {G.}~\bibnamefont {Laskaris}}, \bibinfo
  {author} {\bibfnamefont {M.}~\bibnamefont {Duer}}, \bibinfo {author}
  {\bibfnamefont {V.}~\bibnamefont {Lenivenko}}, \bibinfo {author}
  {\bibfnamefont {E.~P.}\ \bibnamefont {Segarra}}, \bibinfo {author}
  {\bibfnamefont {T.}~\bibnamefont {Atovullaev}}, \bibinfo {author}
  {\bibfnamefont {G.}~\bibnamefont {Johansson}}, \bibinfo {author}
  {\bibfnamefont {T.}~\bibnamefont {Aumann}}, \bibinfo {author} {\bibfnamefont
  {A.}~\bibnamefont {Corsi}},\ and\ \bibinfo {author} {\bibnamefont {et~al.}},\
  }\bibfield  {title} {\bibinfo {title} {Unperturbed inverse kinematics nucleon
  knockout measurements with a carbon beam},\ }\href
  {https://doi.org/10.1038/s41567-021-01193-4} {\bibfield  {journal} {\bibinfo
  {journal} {Nature Physics}\ }\textbf {\bibinfo {volume} {17}},\ \bibinfo
  {pages} {693–699} (\bibinfo {year} {2021})}\BibitemShut {NoStop}%
\bibitem [{\citenamefont {Tang}\ \emph {et~al.}(2003)\citenamefont {Tang} \emph
  {et~al.}}]{tang03}%
  \BibitemOpen
  \bibfield  {author} {\bibinfo {author} {\bibfnamefont {A.}~\bibnamefont
  {Tang}} \emph {et~al.},\ }\bibfield  {title} {\bibinfo {title} {{n-p short
  range correlations from (p,2p + n) measurements}},\ }\href
  {https://doi.org/10.1103/PhysRevLett.90.042301} {\bibfield  {journal}
  {\bibinfo  {journal} {Phys. Rev. Lett.}\ }\textbf {\bibinfo {volume} {90}},\
  \bibinfo {pages} {042301} (\bibinfo {year} {2003})},\ \Eprint
  {https://arxiv.org/abs/nucl-ex/0206003} {arXiv:nucl-ex/0206003} \BibitemShut
  {NoStop}%
\bibitem [{\citenamefont {Piasetzky}\ \emph {et~al.}(2006)\citenamefont
  {Piasetzky}, \citenamefont {Sargsian}, \citenamefont {Frankfurt},
  \citenamefont {Strikman},\ and\ \citenamefont {Watson}}]{piasetzky06}%
  \BibitemOpen
  \bibfield  {author} {\bibinfo {author} {\bibfnamefont {E.}~\bibnamefont
  {Piasetzky}}, \bibinfo {author} {\bibfnamefont {M.}~\bibnamefont {Sargsian}},
  \bibinfo {author} {\bibfnamefont {L.}~\bibnamefont {Frankfurt}}, \bibinfo
  {author} {\bibfnamefont {M.}~\bibnamefont {Strikman}},\ and\ \bibinfo
  {author} {\bibfnamefont {J.~W.}\ \bibnamefont {Watson}},\ }\bibfield  {title}
  {\bibinfo {title} {Evidence for strong dominance of proton-neutron
  correlations in nuclei},\ }\href
  {https://doi.org/10.1103/PhysRevLett.97.162504} {\bibfield  {journal}
  {\bibinfo  {journal} {Phys. Rev. Lett.}\ }\textbf {\bibinfo {volume} {97}},\
  \bibinfo {pages} {162504} (\bibinfo {year} {2006})}\BibitemShut {NoStop}%
\bibitem [{\citenamefont {Shneor}\ \emph {et~al.}(2007)\citenamefont {Shneor}
  \emph {et~al.}}]{shneor07}%
  \BibitemOpen
  \bibfield  {author} {\bibinfo {author} {\bibfnamefont {R.}~\bibnamefont
  {Shneor}} \emph {et~al.} (\bibinfo {collaboration} {Jefferson Lab Hall A}),\
  }\bibfield  {title} {\bibinfo {title} {{Investigation of proton-proton
  short-range correlations via the C-12(e, e-prime pp) reaction}},\ }\href
  {https://doi.org/10.1103/PhysRevLett.99.072501} {\bibfield  {journal}
  {\bibinfo  {journal} {Phys. Rev. Lett.}\ }\textbf {\bibinfo {volume} {99}},\
  \bibinfo {pages} {072501} (\bibinfo {year} {2007})},\ \Eprint
  {https://arxiv.org/abs/nucl-ex/0703023} {arXiv:nucl-ex/0703023} \BibitemShut
  {NoStop}%
\bibitem [{\citenamefont {Subedi}\ \emph {et~al.}(2008)\citenamefont {Subedi}
  \emph {et~al.}}]{subedi08}%
  \BibitemOpen
  \bibfield  {author} {\bibinfo {author} {\bibfnamefont {R.}~\bibnamefont
  {Subedi}} \emph {et~al.},\ }\bibfield  {title} {\bibinfo {title} {{Probing
  Cold Dense Nuclear Matter}},\ }\href
  {https://doi.org/10.1126/science.1156675} {\bibfield  {journal} {\bibinfo
  {journal} {Science}\ }\textbf {\bibinfo {volume} {320}},\ \bibinfo {pages}
  {1476} (\bibinfo {year} {2008})},\ \Eprint {https://arxiv.org/abs/0908.1514}
  {arXiv:0908.1514 [nucl-ex]} \BibitemShut {NoStop}%
\bibitem [{\citenamefont {Korover}\ \emph {et~al.}(2014)\citenamefont
  {Korover}, \citenamefont {Muangma}, \citenamefont {Hen} \emph
  {et~al.}}]{korover14}%
  \BibitemOpen
  \bibfield  {author} {\bibinfo {author} {\bibfnamefont {I.}~\bibnamefont
  {Korover}}, \bibinfo {author} {\bibfnamefont {N.}~\bibnamefont {Muangma}},
  \bibinfo {author} {\bibfnamefont {O.}~\bibnamefont {Hen}}, \emph {et~al.},\
  }\bibfield  {title} {\bibinfo {title} {{Probing the Repulsive Core of the
  Nucleon-Nucleon Interaction via the 4He(e,e'pN) Triple-Coincidence
  Reaction}},\ }\href {https://doi.org/10.1103/PhysRevLett.113.022501}
  {\bibfield  {journal} {\bibinfo  {journal} {Phys. Rev. Lett.}\ }\textbf
  {\bibinfo {volume} {113}},\ \bibinfo {pages} {022501} (\bibinfo {year}
  {2014})}\BibitemShut {NoStop}%
\bibitem [{\citenamefont {Cohen}\ \emph {et~al.}(2018)\citenamefont {Cohen}
  \emph {et~al.}}]{Cohen:2018gzh}%
  \BibitemOpen
  \bibfield  {author} {\bibinfo {author} {\bibfnamefont {E.~O.}\ \bibnamefont
  {Cohen}} \emph {et~al.} (\bibinfo {collaboration} {CLAS Collaboration}),\
  }\bibfield  {title} {\bibinfo {title} {{Center of Mass Motion of Short-Range
  Correlated Nucleon Pairs studied via the $A(e,e'pp)$ Reaction}},\ }\href
  {https://doi.org/10.1103/PhysRevLett.121.092501} {\bibfield  {journal}
  {\bibinfo  {journal} {Phys. Rev. Lett.}\ }\textbf {\bibinfo {volume} {121}},\
  \bibinfo {pages} {092501} (\bibinfo {year} {2018})},\ \Eprint
  {https://arxiv.org/abs/1805.01981} {arXiv:1805.01981 [nucl-ex]} \BibitemShut
  {NoStop}%
\bibitem [{\citenamefont {Hen}\ \emph {et~al.}(2014)\citenamefont {Hen} \emph
  {et~al.}}]{hen14}%
  \BibitemOpen
  \bibfield  {author} {\bibinfo {author} {\bibfnamefont {O.}~\bibnamefont
  {Hen}} \emph {et~al.},\ }\bibfield  {title} {\bibinfo {title} {{Momentum
  sharing in imbalanced Fermi systems}},\ }\href
  {https://doi.org/10.1126/science.1256785} {\bibfield  {journal} {\bibinfo
  {journal} {Science}\ }\textbf {\bibinfo {volume} {346}},\ \bibinfo {pages}
  {614} (\bibinfo {year} {2014})},\ \Eprint {https://arxiv.org/abs/1412.0138}
  {arXiv:1412.0138 [nucl-ex]} \BibitemShut {NoStop}%
\bibitem [{\citenamefont {Duer}\ \emph {et~al.}(2019)\citenamefont {Duer} \emph
  {et~al.}}]{Duer:2018sxh}%
  \BibitemOpen
  \bibfield  {author} {\bibinfo {author} {\bibfnamefont {M.}~\bibnamefont
  {Duer}} \emph {et~al.} (\bibinfo {collaboration} {CLAS Collaboration}),\
  }\bibfield  {title} {\bibinfo {title} {{Direct Observation of Proton-Neutron
  Short-Range Correlation Dominance in Heavy Nuclei}},\ }\href
  {https://doi.org/10.1103/PhysRevLett.122.172502} {\bibfield  {journal}
  {\bibinfo  {journal} {Phys. Rev. Lett.}\ }\textbf {\bibinfo {volume} {122}},\
  \bibinfo {pages} {172502} (\bibinfo {year} {2019})},\ \Eprint
  {https://arxiv.org/abs/1810.05343} {arXiv:1810.05343 [nucl-ex]} \BibitemShut
  {NoStop}%
\bibitem [{\citenamefont {Korover}\ \emph {et~al.}(2021)\citenamefont {Korover}
  \emph {et~al.}}]{CLAS:2020rue}%
  \BibitemOpen
  \bibfield  {author} {\bibinfo {author} {\bibfnamefont {I.}~\bibnamefont
  {Korover}} \emph {et~al.} (\bibinfo {collaboration} {CLAS}),\ }\bibfield
  {title} {\bibinfo {title} {{12C(e,e'pN) measurements of short range
  correlations in the tensor-to-scalar interaction transition region}},\ }\href
  {https://doi.org/10.1016/j.physletb.2021.136523} {\bibfield  {journal}
  {\bibinfo  {journal} {Phys. Lett. B}\ }\textbf {\bibinfo {volume} {820}},\
  \bibinfo {pages} {136523} (\bibinfo {year} {2021})},\ \Eprint
  {https://arxiv.org/abs/2004.07304} {arXiv:2004.07304 [nucl-ex]} \BibitemShut
  {NoStop}%
\bibitem [{\citenamefont {Schmidt}\ \emph {et~al.}(2020)\citenamefont {Schmidt}
  \emph {et~al.}}]{schmidt20}%
  \BibitemOpen
  \bibfield  {author} {\bibinfo {author} {\bibfnamefont {A.}~\bibnamefont
  {Schmidt}} \emph {et~al.} (\bibinfo {collaboration} {CLAS}),\ }\bibfield
  {title} {\bibinfo {title} {{Probing the core of the strong nuclear
  interaction}},\ }\href {https://doi.org/10.1038/s41586-020-2021-6} {\bibfield
   {journal} {\bibinfo  {journal} {Nature}\ }\textbf {\bibinfo {volume}
  {578}},\ \bibinfo {pages} {540} (\bibinfo {year} {2020})},\ \Eprint
  {https://arxiv.org/abs/2004.11221} {arXiv:2004.11221 [nucl-ex]} \BibitemShut
  {NoStop}%
\bibitem [{\citenamefont {Alvioli}\ \emph
  {et~al.}(2013{\natexlab{b}})\citenamefont {Alvioli}, \citenamefont {Ciofi
  Degli~Atti}, \citenamefont {Kaptari}, \citenamefont {Mezzetti},\ and\
  \citenamefont {Morita}}]{Alvioli:2013qyz}%
  \BibitemOpen
  \bibfield  {author} {\bibinfo {author} {\bibfnamefont {M.}~\bibnamefont
  {Alvioli}}, \bibinfo {author} {\bibfnamefont {C.}~\bibnamefont {Ciofi
  Degli~Atti}}, \bibinfo {author} {\bibfnamefont {L.~P.}\ \bibnamefont
  {Kaptari}}, \bibinfo {author} {\bibfnamefont {C.~B.}\ \bibnamefont
  {Mezzetti}},\ and\ \bibinfo {author} {\bibfnamefont {H.}~\bibnamefont
  {Morita}},\ }\bibfield  {title} {\bibinfo {title} {{Universality of
  nucleon-nucleon short-range correlations and nucleon momentum
  distributions}},\ }\href {https://doi.org/10.1142/S021830131330021X}
  {\bibfield  {journal} {\bibinfo  {journal} {Int. J. Mod. Phys.}\ }\textbf
  {\bibinfo {volume} {E22}},\ \bibinfo {pages} {1330021} (\bibinfo {year}
  {2013}{\natexlab{b}})},\ \Eprint {https://arxiv.org/abs/1306.6235}
  {arXiv:1306.6235 [nucl-th]} \BibitemShut {NoStop}%
\bibitem [{\citenamefont {Rios}\ \emph {et~al.}(2014)\citenamefont {Rios},
  \citenamefont {Polls},\ and\ \citenamefont {Dickhoff}}]{Rios:2013zqa}%
  \BibitemOpen
  \bibfield  {author} {\bibinfo {author} {\bibfnamefont {A.}~\bibnamefont
  {Rios}}, \bibinfo {author} {\bibfnamefont {A.}~\bibnamefont {Polls}},\ and\
  \bibinfo {author} {\bibfnamefont {W.~H.}\ \bibnamefont {Dickhoff}},\
  }\bibfield  {title} {\bibinfo {title} {{Density and isospin asymmetry
  dependence of high-momentum components}},\ }\href
  {https://doi.org/10.1103/PhysRevC.89.044303} {\bibfield  {journal} {\bibinfo
  {journal} {Phys. Rev.}\ }\textbf {\bibinfo {volume} {C89}},\ \bibinfo {pages}
  {044303} (\bibinfo {year} {2014})},\ \Eprint
  {https://arxiv.org/abs/1312.7307} {arXiv:1312.7307 [nucl-th]} \BibitemShut
  {NoStop}%
\bibitem [{\citenamefont {Ryckebusch}\ \emph {et~al.}(2019)\citenamefont
  {Ryckebusch}, \citenamefont {Cosyn}, \citenamefont {Vieijra},\ and\
  \citenamefont {Casert}}]{Ryckebusch:2019oya}%
  \BibitemOpen
  \bibfield  {author} {\bibinfo {author} {\bibfnamefont {J.}~\bibnamefont
  {Ryckebusch}}, \bibinfo {author} {\bibfnamefont {W.}~\bibnamefont {Cosyn}},
  \bibinfo {author} {\bibfnamefont {T.}~\bibnamefont {Vieijra}},\ and\ \bibinfo
  {author} {\bibfnamefont {C.}~\bibnamefont {Casert}},\ }\bibfield  {title}
  {\bibinfo {title} {{Isospin composition of the high-momentum fluctuations in
  nuclei from asymptotic momentum distributions}},\ }\href
  {https://doi.org/10.1103/PhysRevC.100.054620} {\bibfield  {journal} {\bibinfo
   {journal} {Phys. Rev. C}\ }\textbf {\bibinfo {volume} {100}},\ \bibinfo
  {pages} {054620} (\bibinfo {year} {2019})},\ \Eprint
  {https://arxiv.org/abs/1907.07259} {arXiv:1907.07259 [nucl-th]} \BibitemShut
  {NoStop}%
\bibitem [{\citenamefont {Neff}\ \emph {et~al.}(2015)\citenamefont {Neff},
  \citenamefont {Feldmeier},\ and\ \citenamefont {Horiuchi}}]{neff15}%
  \BibitemOpen
  \bibfield  {author} {\bibinfo {author} {\bibfnamefont {T.}~\bibnamefont
  {Neff}}, \bibinfo {author} {\bibfnamefont {H.}~\bibnamefont {Feldmeier}},\
  and\ \bibinfo {author} {\bibfnamefont {W.}~\bibnamefont {Horiuchi}},\
  }\bibfield  {title} {\bibinfo {title} {Short-range correlations in nuclei
  with similarity renormalization group transformations},\ }\href
  {https://doi.org/10.1103/PhysRevC.92.024003} {\bibfield  {journal} {\bibinfo
  {journal} {Phys. Rev. C}\ }\textbf {\bibinfo {volume} {92}},\ \bibinfo
  {pages} {024003} (\bibinfo {year} {2015})}\BibitemShut {NoStop}%
\bibitem [{\citenamefont {Ryckebusch}\ \emph {et~al.}(2015)\citenamefont
  {Ryckebusch}, \citenamefont {Vanhalst},\ and\ \citenamefont
  {Cosyn}}]{ryckebusch15}%
  \BibitemOpen
  \bibfield  {author} {\bibinfo {author} {\bibfnamefont {J.}~\bibnamefont
  {Ryckebusch}}, \bibinfo {author} {\bibfnamefont {M.}~\bibnamefont
  {Vanhalst}},\ and\ \bibinfo {author} {\bibfnamefont {W.}~\bibnamefont
  {Cosyn}},\ }\bibfield  {title} {\bibinfo {title} {Stylized features of
  single-nucleon momentum distributions},\ }\href@noop {} {\bibfield  {journal}
  {\bibinfo  {journal} {Journal of Physics G: Nuclear and Particle Physics}\
  }\textbf {\bibinfo {volume} {42}},\ \bibinfo {pages} {055104} (\bibinfo
  {year} {2015})}\BibitemShut {NoStop}%
\bibitem [{\citenamefont {Ciofi~degli Atti}\ \emph {et~al.}(2007)\citenamefont
  {Ciofi~degli Atti}, \citenamefont {Frankfurt}, \citenamefont {Kaptari},\ and\
  \citenamefont {Strikman}}]{CiofidegliAtti:2007ork}%
  \BibitemOpen
  \bibfield  {author} {\bibinfo {author} {\bibfnamefont {C.}~\bibnamefont
  {Ciofi~degli Atti}}, \bibinfo {author} {\bibfnamefont {L.~L.}\ \bibnamefont
  {Frankfurt}}, \bibinfo {author} {\bibfnamefont {L.~P.}\ \bibnamefont
  {Kaptari}},\ and\ \bibinfo {author} {\bibfnamefont {M.~I.}\ \bibnamefont
  {Strikman}},\ }\bibfield  {title} {\bibinfo {title} {{On the dependence of
  the wave function of a bound nucleon on its momentum and the EMC effect}},\
  }\href {https://doi.org/10.1103/PhysRevC.76.055206} {\bibfield  {journal}
  {\bibinfo  {journal} {Phys. Rev.}\ }\textbf {\bibinfo {volume} {C76}},\
  \bibinfo {pages} {055206} (\bibinfo {year} {2007})},\ \Eprint
  {https://arxiv.org/abs/0706.2937} {arXiv:0706.2937 [nucl-th]} \BibitemShut
  {NoStop}%
\bibitem [{\citenamefont {Weinstein}\ \emph {et~al.}(2011)\citenamefont
  {Weinstein}, \citenamefont {Piasetzky}, \citenamefont {Higinbotham},
  \citenamefont {Gomez}, \citenamefont {Hen},\ and\ \citenamefont
  {Shneor}}]{weinstein11}%
  \BibitemOpen
  \bibfield  {author} {\bibinfo {author} {\bibfnamefont {L.~B.}\ \bibnamefont
  {Weinstein}}, \bibinfo {author} {\bibfnamefont {E.}~\bibnamefont
  {Piasetzky}}, \bibinfo {author} {\bibfnamefont {D.~W.}\ \bibnamefont
  {Higinbotham}}, \bibinfo {author} {\bibfnamefont {J.}~\bibnamefont {Gomez}},
  \bibinfo {author} {\bibfnamefont {O.}~\bibnamefont {Hen}},\ and\ \bibinfo
  {author} {\bibfnamefont {R.}~\bibnamefont {Shneor}},\ }\bibfield  {title}
  {\bibinfo {title} {Short range correlations and the emc effect},\ }\href
  {https://doi.org/10.1103/PhysRevLett.106.052301} {\bibfield  {journal}
  {\bibinfo  {journal} {Phys. Rev. Lett.}\ }\textbf {\bibinfo {volume} {106}},\
  \bibinfo {pages} {052301} (\bibinfo {year} {2011})}\BibitemShut {NoStop}%
\bibitem [{\citenamefont {Hen}\ \emph {et~al.}(2012)\citenamefont {Hen},
  \citenamefont {Piasetzky},\ and\ \citenamefont {Weinstein}}]{Hen12}%
  \BibitemOpen
  \bibfield  {author} {\bibinfo {author} {\bibfnamefont {O.}~\bibnamefont
  {Hen}}, \bibinfo {author} {\bibfnamefont {E.}~\bibnamefont {Piasetzky}},\
  and\ \bibinfo {author} {\bibfnamefont {L.~B.}\ \bibnamefont {Weinstein}},\
  }\bibfield  {title} {\bibinfo {title} {New data strengthen the connection
  between short range correlations and the emc effect},\ }\href
  {https://doi.org/10.1103/PhysRevC.85.047301} {\bibfield  {journal} {\bibinfo
  {journal} {Phys. Rev. C}\ }\textbf {\bibinfo {volume} {85}},\ \bibinfo
  {pages} {047301} (\bibinfo {year} {2012})}\BibitemShut {NoStop}%
\bibitem [{\citenamefont {Arrington}\ \emph {et~al.}(2012)\citenamefont
  {Arrington}, \citenamefont {Daniel}, \citenamefont {Day}, \citenamefont
  {Fomin}, \citenamefont {Gaskell},\ and\ \citenamefont
  {Solvignon}}]{Arrington:2012ax}%
  \BibitemOpen
  \bibfield  {author} {\bibinfo {author} {\bibfnamefont {J.}~\bibnamefont
  {Arrington}}, \bibinfo {author} {\bibfnamefont {A.}~\bibnamefont {Daniel}},
  \bibinfo {author} {\bibfnamefont {D.}~\bibnamefont {Day}}, \bibinfo {author}
  {\bibfnamefont {N.}~\bibnamefont {Fomin}}, \bibinfo {author} {\bibfnamefont
  {D.}~\bibnamefont {Gaskell}},\ and\ \bibinfo {author} {\bibfnamefont
  {P.}~\bibnamefont {Solvignon}},\ }\bibfield  {title} {\bibinfo {title} {{A
  detailed study of the nuclear dependence of the EMC effect and short-range
  correlations}},\ }\href {https://doi.org/10.1103/PhysRevC.86.065204}
  {\bibfield  {journal} {\bibinfo  {journal} {Phys. Rev. C}\ }\textbf {\bibinfo
  {volume} {86}},\ \bibinfo {pages} {065204} (\bibinfo {year} {2012})},\
  \Eprint {https://arxiv.org/abs/1206.6343} {arXiv:1206.6343 [nucl-ex]}
  \BibitemShut {NoStop}%
\bibitem [{\citenamefont {Hen}\ \emph {et~al.}(2013)\citenamefont {Hen},
  \citenamefont {Higinbotham}, \citenamefont {Miller}, \citenamefont
  {Piasetzky},\ and\ \citenamefont {Weinstein}}]{Hen:2013oha}%
  \BibitemOpen
  \bibfield  {author} {\bibinfo {author} {\bibfnamefont {O.}~\bibnamefont
  {Hen}}, \bibinfo {author} {\bibfnamefont {D.~W.}\ \bibnamefont
  {Higinbotham}}, \bibinfo {author} {\bibfnamefont {G.~A.}\ \bibnamefont
  {Miller}}, \bibinfo {author} {\bibfnamefont {E.}~\bibnamefont {Piasetzky}},\
  and\ \bibinfo {author} {\bibfnamefont {L.~B.}\ \bibnamefont {Weinstein}},\
  }\bibfield  {title} {\bibinfo {title} {{The EMC Effect and High Momentum
  Nucleons in Nuclei}},\ }\href {https://doi.org/10.1142/S0218301313300178}
  {\bibfield  {journal} {\bibinfo  {journal} {Int. J. Mod. Phys.}\ }\textbf
  {\bibinfo {volume} {E22}},\ \bibinfo {pages} {1330017} (\bibinfo {year}
  {2013})},\ \Eprint {https://arxiv.org/abs/1304.2813} {arXiv:1304.2813
  [nucl-th]} \BibitemShut {NoStop}%
\bibitem [{\citenamefont {Chen}\ \emph {et~al.}(2017)\citenamefont {Chen},
  \citenamefont {Detmold}, \citenamefont {Lynn},\ and\ \citenamefont
  {Schwenk}}]{Chen:2016bde}%
  \BibitemOpen
  \bibfield  {author} {\bibinfo {author} {\bibfnamefont {J.-W.}\ \bibnamefont
  {Chen}}, \bibinfo {author} {\bibfnamefont {W.}~\bibnamefont {Detmold}},
  \bibinfo {author} {\bibfnamefont {J.~E.}\ \bibnamefont {Lynn}},\ and\
  \bibinfo {author} {\bibfnamefont {A.}~\bibnamefont {Schwenk}},\ }\bibfield
  {title} {\bibinfo {title} {{Short Range Correlations and the EMC Effect in
  Effective Field Theory}},\ }\href
  {https://doi.org/10.1103/PhysRevLett.119.262502} {\bibfield  {journal}
  {\bibinfo  {journal} {Phys. Rev. Lett.}\ }\textbf {\bibinfo {volume} {119}},\
  \bibinfo {pages} {262502} (\bibinfo {year} {2017})},\ \Eprint
  {https://arxiv.org/abs/1607.03065} {arXiv:1607.03065 [hep-ph]} \BibitemShut
  {NoStop}%
\bibitem [{\citenamefont {Arrington}\ \emph {et~al.}(2006)\citenamefont
  {Arrington}, \citenamefont {Day}, \citenamefont {Fomin},\ and\ \citenamefont
  {Solvignon-Slifer}}]{Arrington_E12-06-105}%
  \BibitemOpen
  \bibfield  {author} {\bibinfo {author} {\bibfnamefont {J.}~\bibnamefont
  {Arrington}}, \bibinfo {author} {\bibfnamefont {D.}~\bibnamefont {Day}},
  \bibinfo {author} {\bibfnamefont {N.}~\bibnamefont {Fomin}},\ and\ \bibinfo
  {author} {\bibfnamefont {P.}~\bibnamefont {Solvignon-Slifer}},\ }\href@noop
  {} {\bibinfo {title} {E12-06-105: Inclusive scattering from nuclei at $x > 1$
  in the quasielastic and deeply inelastic regimes}} (\bibinfo {year}
  {2006})\BibitemShut {NoStop}%
\bibitem [{\citenamefont {Ye}\ \emph {et~al.}(2018)\citenamefont {Ye} \emph
  {et~al.}}]{Ye18}%
  \BibitemOpen
  \bibfield  {author} {\bibinfo {author} {\bibfnamefont {Z.}~\bibnamefont {Ye}}
  \emph {et~al.} (\bibinfo {collaboration} {The Jefferson Lab Hall A
  Collaboration}),\ }\bibfield  {title} {\bibinfo {title} {Search for
  three-nucleon short-range correlations in light nuclei},\ }\href
  {https://doi.org/10.1103/PhysRevC.97.065204} {\bibfield  {journal} {\bibinfo
  {journal} {Phys. Rev. C}\ }\textbf {\bibinfo {volume} {97}},\ \bibinfo
  {pages} {065204} (\bibinfo {year} {2018})}\BibitemShut {NoStop}%
\bibitem [{\citenamefont {Sargsian}\ \emph {et~al.}(2019)\citenamefont
  {Sargsian}, \citenamefont {Day}, \citenamefont {Frankfurt},\ and\
  \citenamefont {Strikman}}]{Sargasian19}%
  \BibitemOpen
  \bibfield  {author} {\bibinfo {author} {\bibfnamefont {M.~M.}\ \bibnamefont
  {Sargsian}}, \bibinfo {author} {\bibfnamefont {D.~B.}\ \bibnamefont {Day}},
  \bibinfo {author} {\bibfnamefont {L.~L.}\ \bibnamefont {Frankfurt}},\ and\
  \bibinfo {author} {\bibfnamefont {M.~I.}\ \bibnamefont {Strikman}},\
  }\bibfield  {title} {\bibinfo {title} {Searching for three-nucleon
  short-range correlations},\ }\href
  {https://doi.org/10.1103/PhysRevC.100.044320} {\bibfield  {journal} {\bibinfo
   {journal} {Phys. Rev. C}\ }\textbf {\bibinfo {volume} {100}},\ \bibinfo
  {pages} {044320} (\bibinfo {year} {2019})}\BibitemShut {NoStop}%
\bibitem [{\citenamefont {Tan}(2008{\natexlab{a}})}]{Tan08a}%
  \BibitemOpen
  \bibfield  {author} {\bibinfo {author} {\bibfnamefont {S.}~\bibnamefont
  {Tan}},\ }\bibfield  {title} {\bibinfo {title} {Energetics of a strongly
  correlated fermi gas},\ }\href@noop {} {\bibfield  {journal} {\bibinfo
  {journal} {Annals of Physics}\ }\textbf {\bibinfo {volume} {323}},\ \bibinfo
  {pages} {2952} (\bibinfo {year} {2008}{\natexlab{a}})}\BibitemShut {NoStop}%
\bibitem [{\citenamefont {Tan}(2008{\natexlab{b}})}]{Tan08b}%
  \BibitemOpen
  \bibfield  {author} {\bibinfo {author} {\bibfnamefont {S.}~\bibnamefont
  {Tan}},\ }\bibfield  {title} {\bibinfo {title} {Large momentum part of a
  strongly correlated fermi gas},\ }\href
  {https://doi.org/http://dx.doi.org/10.1016/j.aop.2008.03.005} {\bibfield
  {journal} {\bibinfo  {journal} {Annals of Physics}\ }\textbf {\bibinfo
  {volume} {323}},\ \bibinfo {pages} {2971} (\bibinfo {year}
  {2008}{\natexlab{b}})}\BibitemShut {NoStop}%
\bibitem [{\citenamefont {Tan}(2008{\natexlab{c}})}]{Tan08c}%
  \BibitemOpen
  \bibfield  {author} {\bibinfo {author} {\bibfnamefont {S.}~\bibnamefont
  {Tan}},\ }\bibfield  {title} {\bibinfo {title} {Generalized virial theorem
  and pressure relation for a strongly correlated fermi gas},\ }\href
  {https://doi.org/http://dx.doi.org/10.1016/j.aop.2008.03.003} {\bibfield
  {journal} {\bibinfo  {journal} {Annals of Physics}\ }\textbf {\bibinfo
  {volume} {323}},\ \bibinfo {pages} {2987} (\bibinfo {year}
  {2008}{\natexlab{c}})}\BibitemShut {NoStop}%
\bibitem [{\citenamefont {Braaten}(2012)}]{Braaten12}%
  \BibitemOpen
  \bibfield  {author} {\bibinfo {author} {\bibfnamefont {E.}~\bibnamefont
  {Braaten}},\ }\bibfield  {title} {\bibinfo {title} {Universal relations for
  fermions with large scattering length},\ }in\ \href@noop {} {\emph {\bibinfo
  {booktitle} {The BCS-BEC Crossover and the Unitary Fermi Gas}}},\ \bibinfo
  {editor} {edited by\ \bibinfo {editor} {\bibfnamefont {W.}~\bibnamefont
  {Zwerger}}}\ (\bibinfo  {publisher} {Springer},\ \bibinfo {address}
  {Berlin},\ \bibinfo {year} {2012})\BibitemShut {NoStop}%
\bibitem [{\citenamefont {Gandolfi}\ \emph {et~al.}(2011)\citenamefont
  {Gandolfi}, \citenamefont {Schmidt},\ and\ \citenamefont
  {Carlson}}]{gandolfi11}%
  \BibitemOpen
  \bibfield  {author} {\bibinfo {author} {\bibfnamefont {S.}~\bibnamefont
  {Gandolfi}}, \bibinfo {author} {\bibfnamefont {K.~E.}\ \bibnamefont
  {Schmidt}},\ and\ \bibinfo {author} {\bibfnamefont {J.}~\bibnamefont
  {Carlson}},\ }\bibfield  {title} {\bibinfo {title} {Bec-bcs crossover and
  universal relations in unitary fermi gases},\ }\href
  {https://doi.org/10.1103/PhysRevA.83.041601} {\bibfield  {journal} {\bibinfo
  {journal} {Phys. Rev. A}\ }\textbf {\bibinfo {volume} {83}},\ \bibinfo
  {pages} {041601} (\bibinfo {year} {2011})}\BibitemShut {NoStop}%
\bibitem [{\citenamefont {Weiss}\ \emph
  {et~al.}(2015{\natexlab{b}})\citenamefont {Weiss}, \citenamefont {Bazak},\
  and\ \citenamefont {Barnea}}]{Weiss14}%
  \BibitemOpen
  \bibfield  {author} {\bibinfo {author} {\bibfnamefont {R.}~\bibnamefont
  {Weiss}}, \bibinfo {author} {\bibfnamefont {B.}~\bibnamefont {Bazak}},\ and\
  \bibinfo {author} {\bibfnamefont {N.}~\bibnamefont {Barnea}},\ }\bibfield
  {title} {\bibinfo {title} {Nuclear neutron-proton contact and the
  photoabsorption cross section},\ }\href
  {https://doi.org/10.1103/PhysRevLett.114.012501} {\bibfield  {journal}
  {\bibinfo  {journal} {Phys. Rev. Lett.}\ }\textbf {\bibinfo {volume} {114}},\
  \bibinfo {pages} {012501} (\bibinfo {year} {2015}{\natexlab{b}})}\BibitemShut
  {NoStop}%
\bibitem [{\citenamefont {Weiss}\ \emph {et~al.}(2017)\citenamefont {Weiss},
  \citenamefont {Pazy},\ and\ \citenamefont {Barnea}}]{Weiss:2016bxw}%
  \BibitemOpen
  \bibfield  {author} {\bibinfo {author} {\bibfnamefont {R.}~\bibnamefont
  {Weiss}}, \bibinfo {author} {\bibfnamefont {E.}~\bibnamefont {Pazy}},\ and\
  \bibinfo {author} {\bibfnamefont {N.}~\bibnamefont {Barnea}},\ }\bibfield
  {title} {\bibinfo {title} {{Short range correlations - The important role of
  few-body dynamics in many-body systems}},\ }\href
  {https://doi.org/10.1007/s00601-016-1165-2} {\bibfield  {journal} {\bibinfo
  {journal} {Few Body Syst.}\ }\textbf {\bibinfo {volume} {58}},\ \bibinfo
  {pages} {9} (\bibinfo {year} {2017})},\ \Eprint
  {https://arxiv.org/abs/1612.01059} {arXiv:1612.01059 [nucl-th]} \BibitemShut
  {NoStop}%
\bibitem [{\citenamefont {Cruz-Torres}\ \emph {et~al.}(2020)\citenamefont
  {Cruz-Torres}, \citenamefont {Lonardoni}, \citenamefont {Weiss},
  \citenamefont {Piarulli}, \citenamefont {Barnea}, \citenamefont
  {Higinbotham}, \citenamefont {Piasetzky}, \citenamefont {Schmidt},
  \citenamefont {Weinstein}, \citenamefont {Wiringa},\ and\ \citenamefont
  {Hen}}]{Cruz-Torres2020}%
  \BibitemOpen
  \bibfield  {author} {\bibinfo {author} {\bibfnamefont {R.}~\bibnamefont
  {Cruz-Torres}}, \bibinfo {author} {\bibfnamefont {D.}~\bibnamefont
  {Lonardoni}}, \bibinfo {author} {\bibfnamefont {R.}~\bibnamefont {Weiss}},
  \bibinfo {author} {\bibfnamefont {M.}~\bibnamefont {Piarulli}}, \bibinfo
  {author} {\bibfnamefont {N.}~\bibnamefont {Barnea}}, \bibinfo {author}
  {\bibfnamefont {D.~W.}\ \bibnamefont {Higinbotham}}, \bibinfo {author}
  {\bibfnamefont {E.}~\bibnamefont {Piasetzky}}, \bibinfo {author}
  {\bibfnamefont {A.}~\bibnamefont {Schmidt}}, \bibinfo {author} {\bibfnamefont
  {L.~B.}\ \bibnamefont {Weinstein}}, \bibinfo {author} {\bibfnamefont {R.~B.}\
  \bibnamefont {Wiringa}},\ and\ \bibinfo {author} {\bibfnamefont
  {O.}~\bibnamefont {Hen}},\ }\bibfield  {title} {\bibinfo {title} {{Many-body
  factorization and position-momentum equivalence of nuclear short-range
  correlations}},\ }\href {https://doi.org/10.1038/s41567-020-01053-7}
  {\bibfield  {journal} {\bibinfo  {journal} {Nature Physics}\ } (\bibinfo
  {year} {2020})},\ \Eprint {https://arxiv.org/abs/1907.03658}
  {arXiv:1907.03658 [nucl-th]} \BibitemShut {NoStop}%
\bibitem [{\citenamefont {Pybus}\ \emph {et~al.}(2020)\citenamefont {Pybus},
  \citenamefont {Korover}, \citenamefont {Weiss}, \citenamefont {Schmidt},
  \citenamefont {Barnea}, \citenamefont {Higinbotham}, \citenamefont
  {Piasetzky}, \citenamefont {Strikman}, \citenamefont {Weinstein},\ and\
  \citenamefont {Hen}}]{Pybus:2020itv}%
  \BibitemOpen
  \bibfield  {author} {\bibinfo {author} {\bibfnamefont {J.}~\bibnamefont
  {Pybus}}, \bibinfo {author} {\bibfnamefont {I.}~\bibnamefont {Korover}},
  \bibinfo {author} {\bibfnamefont {R.}~\bibnamefont {Weiss}}, \bibinfo
  {author} {\bibfnamefont {A.}~\bibnamefont {Schmidt}}, \bibinfo {author}
  {\bibfnamefont {N.}~\bibnamefont {Barnea}}, \bibinfo {author} {\bibfnamefont
  {D.}~\bibnamefont {Higinbotham}}, \bibinfo {author} {\bibfnamefont
  {E.}~\bibnamefont {Piasetzky}}, \bibinfo {author} {\bibfnamefont
  {M.}~\bibnamefont {Strikman}}, \bibinfo {author} {\bibfnamefont
  {L.}~\bibnamefont {Weinstein}},\ and\ \bibinfo {author} {\bibfnamefont
  {O.}~\bibnamefont {Hen}},\ }\bibfield  {title} {\bibinfo {title}
  {{Generalized contact formalism analysis of the $^4$He$(e,e'pN)$ reaction}},\
  }\href {https://doi.org/10.1016/j.physletb.2020.135429} {\bibfield  {journal}
  {\bibinfo  {journal} {Phys. Lett. B}\ }\textbf {\bibinfo {volume} {805}},\
  \bibinfo {pages} {135429} (\bibinfo {year} {2020})},\ \Eprint
  {https://arxiv.org/abs/2003.02318} {arXiv:2003.02318 [nucl-th]} \BibitemShut
  {NoStop}%
\bibitem [{\citenamefont {Weiss}\ \emph {et~al.}(2021)\citenamefont {Weiss},
  \citenamefont {Denniston}, \citenamefont {Pybus}, \citenamefont {Hen},
  \citenamefont {Piasetzky}, \citenamefont {Schmidt}, \citenamefont
  {Weinstein},\ and\ \citenamefont {Barnea}}]{weiss2020inclusive}%
  \BibitemOpen
  \bibfield  {author} {\bibinfo {author} {\bibfnamefont {R.}~\bibnamefont
  {Weiss}}, \bibinfo {author} {\bibfnamefont {A.~W.}\ \bibnamefont
  {Denniston}}, \bibinfo {author} {\bibfnamefont {J.~R.}\ \bibnamefont
  {Pybus}}, \bibinfo {author} {\bibfnamefont {O.}~\bibnamefont {Hen}}, \bibinfo
  {author} {\bibfnamefont {E.}~\bibnamefont {Piasetzky}}, \bibinfo {author}
  {\bibfnamefont {A.}~\bibnamefont {Schmidt}}, \bibinfo {author} {\bibfnamefont
  {L.~B.}\ \bibnamefont {Weinstein}},\ and\ \bibinfo {author} {\bibfnamefont
  {N.}~\bibnamefont {Barnea}},\ }\bibfield  {title} {\bibinfo {title}
  {Extracting the number of short-range correlated nucleon pairs from inclusive
  electron scattering data},\ }\href
  {https://doi.org/10.1103/PhysRevC.103.L031301} {\bibfield  {journal}
  {\bibinfo  {journal} {Phys. Rev. C}\ }\textbf {\bibinfo {volume} {103}},\
  \bibinfo {pages} {L031301} (\bibinfo {year} {2021})}\BibitemShut {NoStop}%
\bibitem [{\citenamefont {Weiss}\ \emph {et~al.}(2016)\citenamefont {Weiss},
  \citenamefont {Bazak},\ and\ \citenamefont {Barnea}}]{Weiss_EPJA16}%
  \BibitemOpen
  \bibfield  {author} {\bibinfo {author} {\bibfnamefont {R.}~\bibnamefont
  {Weiss}}, \bibinfo {author} {\bibfnamefont {B.}~\bibnamefont {Bazak}},\ and\
  \bibinfo {author} {\bibfnamefont {N.}~\bibnamefont {Barnea}},\ }\bibfield
  {title} {\bibinfo {title} {The generalized nuclear contact and its
  application to the photoabsorption cross-section},\ }\bibfield  {journal}
  {\bibinfo  {journal} {The European Physical Journal A}\ }\textbf {\bibinfo
  {volume} {52}},\ \href {https://doi.org/10.1140/epja/i2016-16092-3}
  {10.1140/epja/i2016-16092-3} (\bibinfo {year} {2016})\BibitemShut {NoStop}%
\bibitem [{\citenamefont {Alvioli}\ \emph {et~al.}(2016)\citenamefont
  {Alvioli}, \citenamefont {Ciofi~degli Atti},\ and\ \citenamefont
  {Morita}}]{Alvioli:2016wwp}%
  \BibitemOpen
  \bibfield  {author} {\bibinfo {author} {\bibfnamefont {M.}~\bibnamefont
  {Alvioli}}, \bibinfo {author} {\bibfnamefont {C.}~\bibnamefont {Ciofi~degli
  Atti}},\ and\ \bibinfo {author} {\bibfnamefont {H.}~\bibnamefont {Morita}},\
  }\bibfield  {title} {\bibinfo {title} {{Universality of nucleon-nucleon
  short-range correlations: the factorization property of the nuclear wave
  function, the relative and center-of-mass momentum distributions, and the
  nuclear contacts}},\ }\href {https://doi.org/10.1103/PhysRevC.94.044309}
  {\bibfield  {journal} {\bibinfo  {journal} {Phys. Rev.}\ }\textbf {\bibinfo
  {volume} {C94}},\ \bibinfo {pages} {044309} (\bibinfo {year}
  {2016})}\BibitemShut {NoStop}%
\bibitem [{\citenamefont {Anderson}\ \emph {et~al.}(2010)\citenamefont
  {Anderson}, \citenamefont {Bogner}, \citenamefont {Furnstahl},\ and\
  \citenamefont {Perry}}]{Anderson2010}%
  \BibitemOpen
  \bibfield  {author} {\bibinfo {author} {\bibfnamefont {E.~R.}\ \bibnamefont
  {Anderson}}, \bibinfo {author} {\bibfnamefont {S.~K.}\ \bibnamefont
  {Bogner}}, \bibinfo {author} {\bibfnamefont {R.~J.}\ \bibnamefont
  {Furnstahl}},\ and\ \bibinfo {author} {\bibfnamefont {R.~J.}\ \bibnamefont
  {Perry}},\ }\bibfield  {title} {\bibinfo {title} {Operator evolution via the
  similarity renormalization group: The deuteron},\ }\href
  {https://doi.org/10.1103/PhysRevC.82.054001} {\bibfield  {journal} {\bibinfo
  {journal} {Phys. Rev. C}\ }\textbf {\bibinfo {volume} {82}},\ \bibinfo
  {pages} {054001} (\bibinfo {year} {2010})}\BibitemShut {NoStop}%
\bibitem [{\citenamefont {Bogner}\ and\ \citenamefont
  {Roscher}(2012)}]{Bogner12}%
  \BibitemOpen
  \bibfield  {author} {\bibinfo {author} {\bibfnamefont {S.~K.}\ \bibnamefont
  {Bogner}}\ and\ \bibinfo {author} {\bibfnamefont {D.}~\bibnamefont
  {Roscher}},\ }\bibfield  {title} {\bibinfo {title} {High-momentum tails from
  low-momentum effective theories},\ }\href
  {https://doi.org/10.1103/PhysRevC.86.064304} {\bibfield  {journal} {\bibinfo
  {journal} {Phys. Rev. C}\ }\textbf {\bibinfo {volume} {86}},\ \bibinfo
  {pages} {064304} (\bibinfo {year} {2012})}\BibitemShut {NoStop}%
\bibitem [{\citenamefont {Tropiano}\ \emph {et~al.}(2021)\citenamefont
  {Tropiano}, \citenamefont {Bogner},\ and\ \citenamefont
  {Furnstahl}}]{Tropiano2021}%
  \BibitemOpen
  \bibfield  {author} {\bibinfo {author} {\bibfnamefont {A.~J.}\ \bibnamefont
  {Tropiano}}, \bibinfo {author} {\bibfnamefont {S.~K.}\ \bibnamefont
  {Bogner}},\ and\ \bibinfo {author} {\bibfnamefont {R.~J.}\ \bibnamefont
  {Furnstahl}},\ }\bibfield  {title} {\bibinfo {title} {Short-range correlation
  physics at low renormalization group resolution},\ }\href
  {https://doi.org/10.1103/PhysRevC.104.034311} {\bibfield  {journal} {\bibinfo
   {journal} {Phys. Rev. C}\ }\textbf {\bibinfo {volume} {104}},\ \bibinfo
  {pages} {034311} (\bibinfo {year} {2021})}\BibitemShut {NoStop}%
\bibitem [{\citenamefont {Beck}\ \emph {et~al.}(2022)\citenamefont {Beck},
  \citenamefont {Weiss},\ and\ \citenamefont {Barnea}}]{Beck_CC}%
  \BibitemOpen
  \bibfield  {author} {\bibinfo {author} {\bibfnamefont {S.}~\bibnamefont
  {Beck}}, \bibinfo {author} {\bibfnamefont {R.}~\bibnamefont {Weiss}},\ and\
  \bibinfo {author} {\bibfnamefont {N.}~\bibnamefont {Barnea}},\ }\href
  {https://doi.org/10.48550/ARXIV.2212.13412} {\bibinfo {title} {On nuclear
  short-range correlations and the zero-energy eigenstates of the schrodinger
  equation}} (\bibinfo {year} {2022})\BibitemShut {NoStop}%
\bibitem [{\citenamefont {Weiss}\ and\ \citenamefont
  {Barnea}(2017)}]{Weiss17_CoupledChannels}%
  \BibitemOpen
  \bibfield  {author} {\bibinfo {author} {\bibfnamefont {R.}~\bibnamefont
  {Weiss}}\ and\ \bibinfo {author} {\bibfnamefont {N.}~\bibnamefont {Barnea}},\
  }\bibfield  {title} {\bibinfo {title} {Contact formalism for coupled
  channels},\ }\href {https://doi.org/10.1103/PhysRevC.96.041303} {\bibfield
  {journal} {\bibinfo  {journal} {Phys. Rev. C}\ }\textbf {\bibinfo {volume}
  {96}},\ \bibinfo {pages} {041303} (\bibinfo {year} {2017})}\BibitemShut
  {NoStop}%
\bibitem [{\citenamefont {Weiss}\ and\ \citenamefont
  {Barnea}(2018)}]{Weiss18_CoupledChannels}%
  \BibitemOpen
  \bibfield  {author} {\bibinfo {author} {\bibfnamefont {R.}~\bibnamefont
  {Weiss}}\ and\ \bibinfo {author} {\bibfnamefont {N.}~\bibnamefont {Barnea}},\
  }\href@noop {} {\bibinfo {title} {The nuclear contact formalism - the
  deuteron channel}} (\bibinfo {year} {2018}),\ \Eprint
  {https://arxiv.org/abs/1801.04526} {arXiv:1801.04526 [nucl-th]} \BibitemShut
  {NoStop}%
\bibitem [{\citenamefont {Alvioli}\ \emph {et~al.}(2008)\citenamefont
  {Alvioli}, \citenamefont {Ciofi~degli Atti},\ and\ \citenamefont
  {Morita}}]{Alvioli:2007zz}%
  \BibitemOpen
  \bibfield  {author} {\bibinfo {author} {\bibfnamefont {M.}~\bibnamefont
  {Alvioli}}, \bibinfo {author} {\bibfnamefont {C.}~\bibnamefont {Ciofi~degli
  Atti}},\ and\ \bibinfo {author} {\bibfnamefont {H.}~\bibnamefont {Morita}},\
  }\bibfield  {title} {\bibinfo {title} {{Proton-neutron and proton-proton
  correlations in medium-weight nuclei and the role of the tensor force}},\
  }\href {https://doi.org/10.1103/PhysRevLett.100.162503} {\bibfield  {journal}
  {\bibinfo  {journal} {Phys. Rev. Lett.}\ }\textbf {\bibinfo {volume} {100}},\
  \bibinfo {pages} {162503} (\bibinfo {year} {2008})}\BibitemShut {NoStop}%
\bibitem [{\citenamefont {Schiavilla}\ \emph {et~al.}(2007)\citenamefont
  {Schiavilla}, \citenamefont {Wiringa}, \citenamefont {Pieper},\ and\
  \citenamefont {Carlson}}]{schiavilla07}%
  \BibitemOpen
  \bibfield  {author} {\bibinfo {author} {\bibfnamefont {R.}~\bibnamefont
  {Schiavilla}}, \bibinfo {author} {\bibfnamefont {R.~B.}\ \bibnamefont
  {Wiringa}}, \bibinfo {author} {\bibfnamefont {S.~C.}\ \bibnamefont
  {Pieper}},\ and\ \bibinfo {author} {\bibfnamefont {J.}~\bibnamefont
  {Carlson}},\ }\bibfield  {title} {\bibinfo {title} {Tensor forces and the
  ground-state structure of nuclei},\ }\href@noop {} {\bibfield  {journal}
  {\bibinfo  {journal} {Phys. Rev. Lett.}\ }\textbf {\bibinfo {volume} {98}},\
  \bibinfo {eid} {132501} (\bibinfo {year} {2007})}\BibitemShut {NoStop}%
\bibitem [{\citenamefont {Braaten}\ \emph {et~al.}(2011)\citenamefont
  {Braaten}, \citenamefont {Kang},\ and\ \citenamefont {Platter}}]{Braaten11}%
  \BibitemOpen
  \bibfield  {author} {\bibinfo {author} {\bibfnamefont {E.}~\bibnamefont
  {Braaten}}, \bibinfo {author} {\bibfnamefont {D.}~\bibnamefont {Kang}},\ and\
  \bibinfo {author} {\bibfnamefont {L.}~\bibnamefont {Platter}},\ }\bibfield
  {title} {\bibinfo {title} {Universal relations for identical bosons from
  three-body physics},\ }\href {https://doi.org/10.1103/PhysRevLett.106.153005}
  {\bibfield  {journal} {\bibinfo  {journal} {Phys. Rev. Lett.}\ }\textbf
  {\bibinfo {volume} {106}},\ \bibinfo {pages} {153005} (\bibinfo {year}
  {2011})}\BibitemShut {NoStop}%
\bibitem [{\citenamefont {Werner}\ and\ \citenamefont
  {Castin}(2012)}]{Werner12_B}%
  \BibitemOpen
  \bibfield  {author} {\bibinfo {author} {\bibfnamefont {F.}~\bibnamefont
  {Werner}}\ and\ \bibinfo {author} {\bibfnamefont {Y.}~\bibnamefont
  {Castin}},\ }\bibfield  {title} {\bibinfo {title} {General relations for
  quantum gases in two and three dimensions. ii. bosons and mixtures},\ }\href
  {https://doi.org/10.1103/PhysRevA.86.053633} {\bibfield  {journal} {\bibinfo
  {journal} {Phys. Rev. A}\ }\textbf {\bibinfo {volume} {86}},\ \bibinfo
  {pages} {053633} (\bibinfo {year} {2012})}\BibitemShut {NoStop}%
\bibitem [{\citenamefont {Bazak}\ \emph {et~al.}(2020)\citenamefont {Bazak},
  \citenamefont {Valiente},\ and\ \citenamefont {Barnea}}]{Bazak20}%
  \BibitemOpen
  \bibfield  {author} {\bibinfo {author} {\bibfnamefont {B.}~\bibnamefont
  {Bazak}}, \bibinfo {author} {\bibfnamefont {M.}~\bibnamefont {Valiente}},\
  and\ \bibinfo {author} {\bibfnamefont {N.}~\bibnamefont {Barnea}},\
  }\bibfield  {title} {\bibinfo {title} {Universal short-range correlations in
  bosonic helium clusters},\ }\href
  {https://doi.org/10.1103/PhysRevA.101.010501} {\bibfield  {journal} {\bibinfo
   {journal} {Phys. Rev. A}\ }\textbf {\bibinfo {volume} {101}},\ \bibinfo
  {pages} {010501} (\bibinfo {year} {2020})}\BibitemShut {NoStop}%
\bibitem [{\citenamefont {Carlson}\ \emph {et~al.}(2015)\citenamefont
  {Carlson}, \citenamefont {Gandolfi}, \citenamefont {Pederiva}, \citenamefont
  {Pieper}, \citenamefont {Schiavilla}, \citenamefont {Schmidt},\ and\
  \citenamefont {Wiringa}}]{Carlson:2014vla}%
  \BibitemOpen
  \bibfield  {author} {\bibinfo {author} {\bibfnamefont {J.}~\bibnamefont
  {Carlson}}, \bibinfo {author} {\bibfnamefont {S.}~\bibnamefont {Gandolfi}},
  \bibinfo {author} {\bibfnamefont {F.}~\bibnamefont {Pederiva}}, \bibinfo
  {author} {\bibfnamefont {S.~C.}\ \bibnamefont {Pieper}}, \bibinfo {author}
  {\bibfnamefont {R.}~\bibnamefont {Schiavilla}}, \bibinfo {author}
  {\bibfnamefont {K.~E.}\ \bibnamefont {Schmidt}},\ and\ \bibinfo {author}
  {\bibfnamefont {R.~B.}\ \bibnamefont {Wiringa}},\ }\bibfield  {title}
  {\bibinfo {title} {{Quantum Monte Carlo methods for nuclear physics}},\
  }\href {https://doi.org/10.1103/RevModPhys.87.1067} {\bibfield  {journal}
  {\bibinfo  {journal} {Rev. Mod. Phys.}\ }\textbf {\bibinfo {volume} {87}},\
  \bibinfo {pages} {1067} (\bibinfo {year} {2015})},\ \Eprint
  {https://arxiv.org/abs/1412.3081} {arXiv:1412.3081 [nucl-th]} \BibitemShut
  {NoStop}%
\bibitem [{\citenamefont {Lonardoni}\ \emph {et~al.}(2018)\citenamefont
  {Lonardoni}, \citenamefont {Gandolfi}, \citenamefont {Lynn}, \citenamefont
  {Petrie}, \citenamefont {Carlson}, \citenamefont {Schmidt},\ and\
  \citenamefont {Schwenk}}]{Lonardoni:2018prc}%
  \BibitemOpen
  \bibfield  {author} {\bibinfo {author} {\bibfnamefont {D.}~\bibnamefont
  {Lonardoni}}, \bibinfo {author} {\bibfnamefont {S.}~\bibnamefont {Gandolfi}},
  \bibinfo {author} {\bibfnamefont {J.~E.}\ \bibnamefont {Lynn}}, \bibinfo
  {author} {\bibfnamefont {C.}~\bibnamefont {Petrie}}, \bibinfo {author}
  {\bibfnamefont {J.}~\bibnamefont {Carlson}}, \bibinfo {author} {\bibfnamefont
  {K.~E.}\ \bibnamefont {Schmidt}},\ and\ \bibinfo {author} {\bibfnamefont
  {A.}~\bibnamefont {Schwenk}},\ }\bibfield  {title} {\bibinfo {title}
  {{Auxiliary field diffusion Monte Carlo calculations of light and medium-mass
  nuclei with local chiral interactions}},\ }\href
  {https://doi.org/10.1103/PhysRevC.97.044318} {\bibfield  {journal} {\bibinfo
  {journal} {Phys. Rev. C}\ }\textbf {\bibinfo {volume} {97}},\ \bibinfo
  {pages} {044318} (\bibinfo {year} {2018})}\BibitemShut {NoStop}%
\bibitem [{\citenamefont {Gezerlis}\ \emph {et~al.}(2013)\citenamefont
  {Gezerlis}, \citenamefont {Tews}, \citenamefont {Epelbaum}, \citenamefont
  {Gandolfi}, \citenamefont {Hebeler}, \citenamefont {Nogga},\ and\
  \citenamefont {Schwenk}}]{Gezerlis:2013ipa}%
  \BibitemOpen
  \bibfield  {author} {\bibinfo {author} {\bibfnamefont {A.}~\bibnamefont
  {Gezerlis}}, \bibinfo {author} {\bibfnamefont {I.}~\bibnamefont {Tews}},
  \bibinfo {author} {\bibfnamefont {E.}~\bibnamefont {Epelbaum}}, \bibinfo
  {author} {\bibfnamefont {S.}~\bibnamefont {Gandolfi}}, \bibinfo {author}
  {\bibfnamefont {K.}~\bibnamefont {Hebeler}}, \bibinfo {author} {\bibfnamefont
  {A.}~\bibnamefont {Nogga}},\ and\ \bibinfo {author} {\bibfnamefont
  {A.}~\bibnamefont {Schwenk}},\ }\bibfield  {title} {\bibinfo {title}
  {{Quantum Monte Carlo Calculations with Chiral Effective Field Theory
  Interactions}},\ }\href {https://doi.org/10.1103/PhysRevLett.111.032501}
  {\bibfield  {journal} {\bibinfo  {journal} {Phys. Rev. Lett.}\ }\textbf
  {\bibinfo {volume} {111}},\ \bibinfo {pages} {032501} (\bibinfo {year}
  {2013})},\ \Eprint {https://arxiv.org/abs/1303.6243} {arXiv:1303.6243
  [nucl-th]} \BibitemShut {NoStop}%
\bibitem [{\citenamefont {Gezerlis}\ \emph {et~al.}(2014)\citenamefont
  {Gezerlis}, \citenamefont {Tews}, \citenamefont {Epelbaum}, \citenamefont
  {Freunek}, \citenamefont {Gandolfi}, \citenamefont {Hebeler}, \citenamefont
  {Nogga},\ and\ \citenamefont {Schwenk}}]{Gezerlis:2014}%
  \BibitemOpen
  \bibfield  {author} {\bibinfo {author} {\bibfnamefont {A.}~\bibnamefont
  {Gezerlis}}, \bibinfo {author} {\bibfnamefont {I.}~\bibnamefont {Tews}},
  \bibinfo {author} {\bibfnamefont {E.}~\bibnamefont {Epelbaum}}, \bibinfo
  {author} {\bibfnamefont {M.}~\bibnamefont {Freunek}}, \bibinfo {author}
  {\bibfnamefont {S.}~\bibnamefont {Gandolfi}}, \bibinfo {author}
  {\bibfnamefont {K.}~\bibnamefont {Hebeler}}, \bibinfo {author} {\bibfnamefont
  {A.}~\bibnamefont {Nogga}},\ and\ \bibinfo {author} {\bibfnamefont
  {A.}~\bibnamefont {Schwenk}},\ }\bibfield  {title} {\bibinfo {title} {{Local
  chiral effective field theory interactions and quantum Monte Carlo
  applications}},\ }\href {https://doi.org/10.1103/PhysRevC.90.054323}
  {\bibfield  {journal} {\bibinfo  {journal} {Phys. Rev. C}\ }\textbf {\bibinfo
  {volume} {90}},\ \bibinfo {pages} {054323} (\bibinfo {year}
  {2014})}\BibitemShut {NoStop}%
\bibitem [{\citenamefont {Lynn}\ \emph {et~al.}(2016)\citenamefont {Lynn},
  \citenamefont {Tews}, \citenamefont {Carlson}, \citenamefont {Gandolfi},
  \citenamefont {Gezerlis}, \citenamefont {Schmidt},\ and\ \citenamefont
  {Schwenk}}]{Lynn:2016}%
  \BibitemOpen
  \bibfield  {author} {\bibinfo {author} {\bibfnamefont {J.~E.}\ \bibnamefont
  {Lynn}}, \bibinfo {author} {\bibfnamefont {I.}~\bibnamefont {Tews}}, \bibinfo
  {author} {\bibfnamefont {J.}~\bibnamefont {Carlson}}, \bibinfo {author}
  {\bibfnamefont {S.}~\bibnamefont {Gandolfi}}, \bibinfo {author}
  {\bibfnamefont {A.}~\bibnamefont {Gezerlis}}, \bibinfo {author}
  {\bibfnamefont {K.~E.}\ \bibnamefont {Schmidt}},\ and\ \bibinfo {author}
  {\bibfnamefont {A.}~\bibnamefont {Schwenk}},\ }\bibfield  {title} {\bibinfo
  {title} {{Chiral Three-Nucleon Interactions in Light Nuclei, Neutron-$\alpha$
  Scattering, and Neutron Matter}},\ }\href
  {https://doi.org/10.1103/PhysRevLett.116.062501} {\bibfield  {journal}
  {\bibinfo  {journal} {Phys. Rev. Lett.}\ }\textbf {\bibinfo {volume} {116}},\
  \bibinfo {pages} {062501} (\bibinfo {year} {2016})}\BibitemShut {NoStop}%
\bibitem [{\citenamefont {Sargsian}(2014)}]{Sargsian:2012sm}%
  \BibitemOpen
  \bibfield  {author} {\bibinfo {author} {\bibfnamefont {M.~M.}\ \bibnamefont
  {Sargsian}},\ }\bibfield  {title} {\bibinfo {title} {{New properties of the
  high-momentum distribution of nucleons in asymmetric nuclei}},\ }\href
  {https://doi.org/10.1103/PhysRevC.89.034305} {\bibfield  {journal} {\bibinfo
  {journal} {Phys. Rev. C}\ }\textbf {\bibinfo {volume} {89}},\ \bibinfo
  {pages} {034305} (\bibinfo {year} {2014})},\ \Eprint
  {https://arxiv.org/abs/1210.3280} {arXiv:1210.3280 [nucl-th]} \BibitemShut
  {NoStop}%
\bibitem [{\citenamefont {Sargsian}()}]{Sargsian_private}%
  \BibitemOpen
  \bibfield  {author} {\bibinfo {author} {\bibfnamefont {M.~M.}\ \bibnamefont
  {Sargsian}},\ }\href@noop {} {}\bibinfo {howpublished} {(private
  communication)}\BibitemShut {NoStop}%
\end{thebibliography}%

\end{document}